\documentclass[aps, prd, twocolumn, 10pt,
               superscriptaddress, nofootinbib, 
               longbibliography, floatfix, 
               preprintnumbers]{revtex4-2}

\usepackage{amsmath,amssymb,bm}
\usepackage{booktabs}
\usepackage{graphicx}
\usepackage{microtype}
\usepackage[dvipsnames]{xcolor}
\usepackage[normalem]{ulem}
\usepackage[colorlinks=true,
            citecolor=MidnightBlue,
            linkcolor=MidnightBlue,
            urlcolor=MidnightBlue]{hyperref}
\hypersetup{breaklinks=true}
\usepackage{orcidlink}
\allowdisplaybreaks

\newcommand{\LCDM}{\ensuremath{\Lambda\mathrm{CDM}}}

\newcommand{\dd}{\mathrm{d}}
\newcommand{\e}{\mathrm{e}}
\newcommand{\Geff}{G_{\mathrm{eff}}}
\newcommand{\GN}{G_{\mathrm N}}
\newcommand{\Pc}{\mathcal{P}}
\newcommand{\Rlag}{R_{\mathrm L}}

\newcommand{\kMpc}{h\,\mathrm{Mpc}^{-1}}
\newcommand{\mgcola}{\texttt{MG-PICOLA}}
\DeclareMathOperator{\Var}{Var}
\newcommand{\Dev}{\mathcal D}
\newcommand{\mDev}{\overline{\mathcal D}}
\newcommand{\Resp}{\mathcal R}
\newcommand{\class}{\texttt{CLASS}}

\begin{document}
\preprint{YITP-26-170}

\title{Phantom memories: how the matter-density PDF records a phantom-crossing expansion history in VCDM}

\author{Antonio De Felice\orcidlink{0000-0002-5556-4693}}
\thanks{The authors after the first two are listed in alphabetical order by surname.}
\affiliation{Center for Gravitational Physics and Quantum Information, Yukawa Institute for Theoretical Physics, Kyoto University, 606-8502 Kyoto, Japan}

\author{Lina Castiblanco\orcidlink{0000-0002-2324-7335}}
\email[Corresponding author: ]{lcastiblancotolosa@physik.uni-bielefeld.de}
\affiliation{Fakultät für Physik, Universität Bielefeld, Universitätsstrasse 25, 33615 Bielefeld, Germany}

\author{Alexander Gallego-Cadavid\orcidlink{0000-0002-9126-9130}}
\author{Pedro M. Ibarbo-Perlaza\orcidlink{0000-0001-7374-7613}}
\author{J. Bayron Orjuela-Quintana\orcidlink{0000-0001-5090-2860}}
\author{Jose L. Palacios-C\'ordoba}
\affiliation{Departamento  de  F\'isica,  Universidad  del Valle, Ciudad  Universitaria Mel\'endez, Santiago de Cali  760032,  Colombia}
\author{Cora Uhlemann\orcidlink{0000-0001-7831-1579}} 
\affiliation{Fakultät für Physik, Universität Bielefeld, Universitätsstrasse 25, 33615 Bielefeld, Germany}

\author{Jack Zidich}
\affiliation{Center for Gravitational Physics and Quantum Information, Yukawa Institute for Theoretical Physics, Kyoto University, 606-8502 Kyoto, Japan}

\date{26 November 2026}

\begin{abstract}
The one-point probability distribution function (PDF) of the matter density in spheres describes how cosmic volume is distributed across density environments. We investigate its response to the phantom-crossing F42 expansion history in VCDM, a theory with a Newtonian matter force at leading subhorizon order. We run three suites of twenty COLA $N$-body simulations with identical initial phases: VCDM with the F42 best-fit parameters from Planck 2018 and DESI DR2 BAO data; a matched $\Lambda$CDM with the same parameters, isolating the effects of the expansion history; and $\Lambda$CDM with its own best-fit parameters. With respect to the matched $\Lambda$CDM model, we find a sign-changing response in the matter power spectrum: it is enhanced by $1.2\%$ at $z=1$, nearly unchanged at $z=0.5$, and suppressed by $1.5\%$ at $z=0$. The slower early expansion of VCDM first enhances the growth of structure, and its faster late expansion then suppresses it. As a result, the width of the PDF follows the same sequence, with a log-density variance about $1\%$ larger than in $\Lambda$CDM at $z=1$ and about $1\%$ smaller at $z=0$. Most of the PDF response is associated with this change in width, while centering and standardizing the log-density leaves much smaller residual shape differences. Theoretical predictions based on large-deviation theory with a simulation-calibrated log-density variance reproduce the shape of the PDF response to both the expansion history and the parameter shift. 
These results show that the nonlinear matter distribution retains an imprint of the accumulated growth history induced by the modified cosmic expansion, even when the short-range gravitational force is unchanged.

\end{abstract}

\maketitle

\section{Introduction}
\label{sec:introduction}

While the standard $\Lambda$-cold dark matter (\LCDM) model successfully describes a broad range of cosmological observations,
persistent theoretical puzzles---such as the cosmological constant problem~\cite{Weinberg:1988cp, Martin:2012bt}---and reported tensions in the Hubble constant $H_0$~\cite{Riess:2021jrx, DiValentino:2021izs, Riess:2024ohe} motivate the exploration of extended dark energy and modified gravity frameworks~\cite{Clifton:2011jh, Koyama:2015vza, CANTATA:2021asi}. In particular, recent DESI BAO measurements, combined with CMB and supernova data, show hints of a dynamical dark energy whose equation of state crosses $w=-1$~\cite{DESI:2025zgx}. At the same time, increasingly precise observations of the large-scale structure (LSS) of the Universe~\cite{LSST:2008ijt, eBOSS:2020yzd, DESI:2024mwx, Euclid:2025rvk} provide new opportunities to test these alternatives through the gravitational clustering of matter~\cite{Joyce:2014kja, Ishak:2018his, Ye:2024ywg, Ishak:2024jhs}.

For Gaussian initial conditions, the linear matter density field is statistically fully characterized by its power spectrum~\cite{Bardeen:1985tr}. As gravitational evolution becomes nonlinear, however, higher-order correlations develop, and the power spectrum alone no longer provides a complete statistical description~\cite{Peebles:1980yev, Fry:1983cj, Gaztanaga:1993ru, Scoccimarro:2000sp, Bernardeau:2002, Szapudi2009}. The one-point probability distribution function (PDF) of the matter density in spheres of comoving radius $R$, closely related to counts-in-cells statistics~\cite{White:1979kp, Bernardeau:1994,
Valageas:2001zr, Ivanov:2018lcg}, provides complementary information~\cite{Uhlemann:2020}. It describes the probability of finding a given normalized cell density $\rho_R=1+\delta_R$, and hence how cosmic volume is distributed among underdense and overdense environments.

A theoretical prediction for this PDF requires following the density field into the nonlinear regime. Large-deviation theory (LDT) provides such a prediction by connecting the final cell density to the initial fluctuations. It associates a final cell density with an initial fluctuation through spherical-collapse dynamics and constructs its rate function from the initial statistics, often for the Gaussian case~\cite{Bernardeau:1994, Valageas:2001zr, Bernardeau:2013dua, Bernardeau:2016}. A saddle-point approximation, conveniently formulated in terms of the log-density, then provides an analytical model for the nonlinear PDF~\cite{Uhlemann:2016}. The framework has also been extended
to evolving dark energy and modified gravity cosmologies~\cite{Gough:2021hlr, Cataneo:2022} and to primordial
non-Gaussianity~\cite{Uhlemann:2017, Friedrich:2020}. In this work, we follow the formalism of Ref.~\cite{Cataneo:2022}, but solve the spherical collapse with the full background of each model rather than in its Einstein--de Sitter approximation.

In most extended models, however, a modified expansion history comes together with a modified gravitational force, and their effects on the matter distribution are mixed. An important question is therefore whether a modified expansion history alone can leave a distinguishable imprint on the matter distribution, when the gravitational force on small scales remains unchanged. VCDM provides a natural framework to
address it. Within the broader class of minimally modified gravity theories ~\cite{Aoki:2018brq, Aoki:2020iwm}, VCDM belongs to the Type-II class and propagates only the two tensor degrees of freedom of General Relativity (GR). Its auxiliary scalar sector modifies the cosmological background without introducing an additional propagating gravitational scalar~\cite{DeFelice:2020typeII, DeFelice:2020cpt, 
DeFelice:2022uxv}. For pressureless matter, the gravitational source approaches its standard Newtonian form in the deep subhorizon limit, while the expansion history and the evolution of perturbations at finite wavelengths can differ from those of \LCDM. 

We focus on the phantom-crossing F42 background~\cite{Borghetto:2026x}, which provides a concrete example of a modified cosmic expansion history without requiring a modified short-range force in the Newtonian approximation adopted here. Importantly, a different expansion history changes not only the instantaneous growth rate but also the accumulated growth of matter fluctuations. The resulting response need not have the same sign at different cosmic epochs.

We study the imprint of the expansion history on the background and linear growth, computed from the model equations, and on the matter power spectrum and one-point PDF, measured in three sets of phase-matched dark matter $N$-body simulations. The best-fit cosmological parameters are first determined independently for F42 and \LCDM{} using Planck 2018 cosmic microwave background (CMB)~\cite{Planck:2018vyg, Planck:2019nip, Planck:2018lbu} and DESI DR2 baryon acoustic oscillation (BAO)~\cite{DESI:2025zgx} data. Using the F42 best-fit parameters, we construct two simulation suites: $V$, with the F42 expansion history, and $\Lambda_V$, with the standard \LCDM{} expansion history but otherwise identical cosmological parameters and initial phases. A third suite, $\Lambda_{\rm fit}$, uses the independently determined \LCDM{} best-fit parameters. Within the adopted Newtonian description, comparing $V$ with $\Lambda_V$ isolates the effect of the expansion history at fixed cosmological parameters and particle force. Comparing $\Lambda_V$ with $\Lambda_{\rm fit}$ quantifies the effect of refitting the parameters, and comparing $V$ with $\Lambda_{\rm fit}$ combines both. We compare the LDT predictions with all three comparisons, and quantify
the agreement for the controlled comparison between $V$ and $\Lambda_V$. For this, we take the nonlinear log-density variance directly from the simulations of each cosmology. LDT then predicts the shape of the PDF response for the measured widths.

Our main results can be summarized as follows. Starting from the same initial fluctuations, the change of expansion history leaves an accumulated imprint on the matter power spectrum whose sign changes with time. This sign change follows the accumulated growth of fluctuations rather than the instantaneous equation of state. It therefore does not occur at the phantom crossing, but later, once the faster late-time expansion has erased the growth advantage that $V$ gained at early times. When $V$ is instead compared with $\Lambda_{\rm fit}$, so that both
models describe the same CMB and BAO data, the difference in the power spectrum is dominated by the shift in the standard parameters, but the effect of the expansion history remains visible on top of it. The PDF follows the same sign-changing sequence, mainly through the change in its width, while the spherical-collapse dynamics barely changes; in the PDF, the expansion history even dominates the difference between the two fitted models at $z=1$. Once the measured log-density variance is supplied, LDT reproduces the PDF responses in all three comparisons.

This work is organized as follows. In Sec.~\ref{sec:f42}, we review VCDM theory, its background dynamics, and the observational constraints derived for F42 and \LCDM. In Sec.~\ref{sec:ldt}, we describe spherical collapse and the LDT construction of the matter-density PDF. Section~\ref{sec:numerics} presents the matched simulation suites and statistical estimators. Our results are presented in Sec.~\ref{sec:results}, followed by a discussion of their physical interpretation and limitations in Sec.~\ref{sec:discussion}. We summarize our conclusions in Sec.~\ref{sec:conclusions}, while numerical consistency and validation checks are collected in the Appendices.

\section{F42: a different clock with a standard subhorizon force}
\label{sec:f42}

We first review the cosmological dynamics relevant for the structure formation analysis, from the underlying gravitational theory to the evolution of matter perturbations. This provides the background and linear quantities required for the nonlinear calculations presented in the following sections.

\subsection{VCDM theory: background and linear growth}
\label{sec:vcdm-theory}

VCDM is a Type-II minimally modified gravity theory in which the cosmological constant of \LCDM{} is replaced by a function $V(\phi)$ of an auxiliary field $\phi$~\cite{DeFelice:2020typeII, DeFelice:2020cpt, DeFelice:2022uxv}. The theory is defined in a preferred time slicing, in which temporal diffeomorphism invariance is broken while spatial diffeomorphisms are preserved, and $\phi$ enters without time derivatives. The constraints of the theory remove the would-be scalar graviton, so that, as in GR, only two tensor polarizations propagate. Although the scalar field $\phi$ is not a quintessence-like field, the function $V(\phi)$ can modify the background expansion and, through the constraints, the evolution of matter perturbations.

In VCDM, the function $V(\phi)$ can be reconstructed from a prescribed expansion history. On a homogeneous Friedmann--Lema\^itre--Robertson--Walker (FLRW) background, the VCDM modification can be written as an effective dark-energy component. Denoting by $\bar{\rho}_{\rm std}$ and $\bar{p}_{\rm std}$ the total background energy density and pressure of the standard matter species, its energy density and pressure are~\cite{DeFelice:2020typeII}:
\begin{align}
 \rho_\phi &\equiv M_{\rm Pl}^2\left(V-\phi V_{,\phi} +\frac34 V_{,\phi}^2\right), \label{eq:vcdm-rhophi} \\
 p_\phi &\equiv -\rho_\phi-\frac32(\bar{\rho}_{\rm std}+\bar{p}_{\rm std})V_{,\phi\phi}. \label{eq:vcdm-pphi}
\end{align}
Here, a comma followed by $\phi$ denotes differentiation with respect to the auxiliary field, and $M_{\rm Pl}^{-2} \equiv 8\pi \GN$. Thus, $w_\phi \equiv p_\phi/\rho_\phi$ is an effective equation of state (EoS) rather than the equation of state of a fundamental fluid. In particular, $w_\phi < - 1$ or a crossing of $w_\phi = -1$ need not imply a ghost-like scalar mode. 

The background reconstruction is especially transparent in terms of the $e$-fold variable $\mathcal N = \ln a$.  The independent equations can be written as \cite{DeFelice:2020typeII}:
\begin{align}
 V(\phi) &= \frac{\phi^2}{3}-\frac{\bar{\rho}_{\rm std}}{M_{\rm Pl}^2}, \label{eq:vcdm-Vrec} \\
 \frac{\dd\phi}{\dd\mathcal N} &=\frac32\frac{\bar{\rho}_{\rm std}+\bar{p}_{\rm std}}{M_{\rm Pl}^2H}, \label{eq:vcdm-phirec}\\
 \frac{\dd\bar{\rho}_\mathrm{std}}{\dd\mathcal N} &=-3(\bar{\rho}_\mathrm{std} + \bar{p}_\mathrm{std}). \label{eq:vcdm-rhoconservation}
\end{align}
Equivalently, away from the special case $\bar{\rho}_{\rm std}+\bar{p}_{\rm std}=0$,
\begin{equation}
 \phi=\frac32 V_{,\phi}-3H.
 \label{eq:vcdm-phiH}
\end{equation}
For an expanding background with $\bar{\rho}_{\rm std}+\bar{p}_{\rm std}>0$, $\phi(\mathcal N)$ is monotonic and can be inverted. A specified expansion history (or ``background clock''), encoded in the Hubble parameter $H$, can therefore be mapped into a corresponding $V(\phi)$.  

We take this to be the one-parameter F42 history of Ref.~\cite{Borghetto:2026x}, with effective dark energy EoS:
\begin{equation}
  w_\phi(a) = \frac{w_0}{\sqrt a},
  \label{eq:f42-w}
\end{equation}
where $a$ is the scale factor and $w_0 \equiv w_\phi(a=1)$. For the best-fit value $w_0\simeq-0.79$ considered below, this effective EoS is phantom at early times and crosses $w_\phi=-1$ at low redshift, reaching $w_\phi=w_0$ today. This regular effective phantom crossing is compatible with VCDM and does not introduce a propagating ghost-like scalar mode. The fixed-exponent F42 family does not reduce to \LCDM{} for any value of $w_0$~\cite{Borghetto:2026x}. 

The relation between this background modification and matter clustering can
be made explicit. For a single pressureless fluid (i.e., neglecting radiation), Ref.~\cite{DeFelice:2020typeII} integrates out all scalar auxiliary variables and obtains
the exact finite-wavelength equation of motion for the gauge-invariant dust contrast, $\delta_m$, which may be written schematically as:
\begin{equation}
 \delta_m''+\mathcal F(\mathcal K,a)\,\delta_m' -\frac32\Omega_m\,\mu_g(\mathcal K,a)\,\delta_m=0,
 \label{eq:dust-eom-schematic}
\end{equation}
where $\mathcal K\equiv k/(aH)$ (with $c=1$), $k$ is the comoving wavenumber, $\Omega_m \equiv \bar\rho_m/(3M_{\rm Pl}^2H^2)$ is the matter density fraction, and a prime denotes differentiation with respect to $\mathcal N$ at fixed $k$. The source coefficient of this reduced single-dust equation is~\cite{DeFelice:2020typeII}:

\begin{equation}
 \mu_g(\mathcal K,a) \equiv \frac{\Geff}{\GN}
 =
 \frac{\mathcal K^2-3H'/H}
 {\mathcal K^2+\frac92\Omega_m} .
 \label{eq:geff-finite-H}
\end{equation}
The friction coefficient follows from the same reduced quadratic action,
\begin{equation}
 \mathcal F(\mathcal K,a) \equiv 2+\frac{H'}{H}-\frac{\mu_g'}{\mu_g},
 \label{eq:friction-exact}
\end{equation}
where the derivative of $\mu_g$ includes the implicit time dependence of $\mathcal K$ at fixed comoving $k$. Thus, at finite wavelength, the background clock affects both the gravitational source and the friction term.
If the same background is represented as dust plus an
effective VCDM dark energy component with density fraction $\Omega_\phi \equiv \rho_\phi/(3 M_\mathrm{Pl}^2 H^2)$, then:
\begin{equation}
 \frac{H'}{H}
 =-\frac32\left[\Omega_m+(1+w_\phi)\Omega_\phi\right] ,
 \label{eq:Hprime-effectiveDE}
\end{equation}
and Eq.~\eqref{eq:geff-finite-H} becomes
\begin{equation}
 \mu_g(\mathcal{K}, a) = 1 + \frac92\frac{(1+w_\phi)\Omega_\phi}{\mathcal K^2+\frac92\Omega_m}.
 \label{eq:geff-finite-w}
\end{equation}
Two complementary GR limits are now manifest. First, at fixed time and in the deep-subhorizon limit, Eqs.~\eqref{eq:geff-finite-w} and \eqref{eq:friction-exact} give
\begin{equation}
 \mu_g=1+\mathcal O(\mathcal K^{-2}),\qquad
 \mathcal F=2+\frac{H'}{H}+\mathcal O(\mathcal K^{-2}),
 \label{eq:gr-subhorizon-limits}
\end{equation}
for a regular dust background. Consistently with the weak-field, short-scale GR limit of VCDM~\cite{DeFelice:2020typeII}, the leading gravitational constraint takes the standard Newtonian Poisson form,
\begin{equation}
 \nabla_{\boldsymbol x}^{2}\Phi_{\rm N}
 \simeq 4\pi\GN a^2\bar\rho_m\,\delta_m,
 \qquad \mathcal K\gg1,
 \label{eq:poisson-subhorizon}
\end{equation}
where $\boldsymbol x$ denotes comoving coordinates and $\Phi_{\rm N}$ the Newtonian potential. Here the equality is understood to leading order in the weak-field, subhorizon expansion, not as an exact finite-wavelength relation. The $N$-body simulations adopt the corresponding standard-force closure for nonlinear matter density contrasts and Newtonian particle trajectories; the linear dust reduction alone does not establish its exact validity at all nonlinear orders.

Second, Ref.~\cite{DeFelice:2020typeII} shows that the reduced scalar action coincides with the GR action when evaluated on the full \LCDM{} background. For late-time dust plus a cosmological constant, this condition is $H'=-\frac32\Omega_mH$, which implies $V_{,\phi\phi}=0$ and makes $\mu_g=1$ at every $\mathcal K$. If this condition is satisfied only at an isolated instant, the source coefficient agrees with GR at that instant, but the full friction coefficient need not, because $\mu_g'$ probes the time evolution of the background. The F42 history thus provides a controlled example in which the cosmological expansion differs from \LCDM{} while the simulated short-scale particle force retains its standard Newtonian form.

\subsection{Radiation-aware background}
\label{sec:background}

Next, we turn to describe the background equations of motion from F42 in the VCDM theory. Integrating the continuity equation, $\rho'_\phi = -3(1 + w_\phi)\rho_\phi$, for $w_\phi$ in Eq.~\eqref{eq:f42-w} gives:
\begin{align}
 \rho_\phi (a) &= \rho_{\phi,0}\,a^{-3} X_\phi(a), \label{eq:f42-rho} \\
 X_\phi(a) &\equiv \exp\!\left[6w_0\left(a^{-1/2}-1\right)\right]. \label{eq:x-def}
\end{align} 
Then, the background histories to be implemented in the numerical pipeline are:
\begin{align}
 E_{\rm F42}^2(a) &= \Omega_{r,0} a^{-4} + a^{-3} \left[\Omega_{m,0} + \Omega_{\phi,0} X_\phi(a)\right], \label{eq:f42-E} \\
 E_{\Lambda{\rm CDM}}^2(a) &= \Omega_{r,0} a^{-4} + \Omega_{m,0} a^{-3} + \Omega_{\Lambda,0}, \label{eq:lcdm-E}
\end{align}
where $E \equiv H/H_0$ and $\Omega_{\phi,0}=\Omega_{\Lambda,0} = 1-\Omega_{m,0}-\Omega_{r,0}$, with the subindex $0$ denoting that the density fractions are evaluated at $z = 0$. Although the contribution of the radiation fluid to the energy budget is small at late-times, retaining it makes the initialization and all numerical inputs internally consistent.

\begin{table*}[t]
 \caption{Cosmological parameters of the three simulation suites, from the
 best fits to Planck 2018 and DESI DR2 BAO data with massless neutrinos
 ($N_{\rm ur}=3.046$, $N_{\rm ncdm}=0$). $V$ uses the F42 best fit, in
 which $w_0$ is the single parameter of the F42 background,
 $w(a)=w_0/\sqrt{a}$ (Sec.~\ref{sec:f42}). $\Lambda_V$ shares all
 parameters of $V$ but has a cosmological constant, and
 $\Lambda_{\rm fit}$ is the independent $\Lambda$CDM best fit. The values
 of $\sigma_8$ are derived from the linear spectra used to initialise the
 simulations. The best fits differ by
 $\Delta\chi^2=\chi^2_{\rm F42}-\chi^2_{\Lambda{\rm CDM}}=-5.29$.
 Full-precision values are given in Appendix~\ref{app:likelihood-breakdown}.}
 \label{tab:cosmology}
 \centering
 \begin{ruledtabular}
 \begin{tabular}{lccccccccc}
 Suite & $H_0$ & $\Omega_m$ & $\Omega_b$ & $10^9A_s$ & $n_s$
       & $\tau_{\rm reio}$ & $w_0$ & $\sigma_8$ & $\chi^2_{\rm opt}$ \\
 \hline
 $V$                 & 66.57 & 0.3202 & 0.05068 & 2.091 & 0.9680 & 0.0528 & $-0.787$ & 0.8107 & 2780.18 \\
 $\Lambda_V$         & 66.57 & 0.3202 & 0.05068 & 2.091 & 0.9680 & 0.0528 & $-1.00$     & 0.8167 & ---     \\
 $\Lambda_{\rm fit}$ & 68.67 & 0.2987 & 0.04773 & 2.103 & 0.9704 & 0.0564 & $-1.00$     & 0.8208 & 2785.48 \\
 \end{tabular}
 \end{ruledtabular}
\end{table*}

\subsection{Massless-neutrino CMB+DESI likelihood records}
\label{sec:fit-check}

To set the parameters of the models (F42 and \LCDM), we use the Planck 2018 high-$\ell$ TTTEEE, low-$\ell$ EE, low-$\ell$ TT, and lensing likelihoods, together with DESI DR2 BAO \cite{Planck:2019nip, Planck:2018lbu, DESI:2025zgx}. In both models, we fix the number of non-CDM species and the number of ultrarelativistic species as $N_{\rm ncdm}=0$ and $N_{\rm ur} = 3.046$, respectively. Therefore, neutrinos are considered massless particles here. Both the production posterior chains and the dedicated \textit{annealed-MCMC minimizations} are run with \texttt{MontePython}~\cite{Audren:2012wb, Brinckmann:2018cvx}\footnote{\url{https://github.com/brinckmann/montepython_public}}. The likelihood minima are obtained by annealed minimization rather than by selecting the best sampled point of an ordinary posterior chain. 

For each model, the converged long-chain covariance and best-fit estimate initialize the minimization. Three successive $10^4$-step stages use likelihood factors $\beta=1,200,1000$ and proposal jumping factors $0.5,0.1,0.05$, respectively. Here $\beta$ multiplies the log likelihood, so that the likelihood-dependent sampling weight is $\mathcal L^\beta=\exp[-\beta\chi^2/2]$; increasing $\beta$ concentrates the search around its maxima. The proposal covariance is held fixed, and each new stage starts from the preceding stage's best point. The final objective is then evaluated with the original, \textit{unscaled} likelihood~(see Table~\ref{tab:likelihood-breakdown} in Appendix~\ref{app:likelihood-breakdown}). Throughout this paper, ``best fit'' or ``likelihood minimum'' denotes the numerical minimum found by this procedure. The production long chains alone provide posterior statistics and posterior plots; the annealed chains are used only for minimization, not as posterior samples. The result of this procedure is shown in Table~\ref{tab:cosmology}.

The phantom crossing follows directly from $w(a_\times)=-1$, \begin{equation}
 a_\times=w_0^2,  \quad  z_\times=w_0^{-2}-1 \approx 0.61,
 \label{eq:crossing}
\end{equation}
where we used the best-fit value $w_0=-0.7873$. 

Table~\ref{tab:cosmology-full} lists the input parameters of the three simulation suites at full precision. The values of $\sigma_8$ are not input parameters but are derived from the linear spectra used to initialise the simulations. They differ between $V$ and $\Lambda_V$, despite the shared parameters, because the two models grow structure differently from the same initial amplitude. 

The difference between the two best fits is
\begin{equation}
 \Delta\chi^2_{\rm F42-\Lambda CDM} \equiv
 \chi^2_{\rm opt,F42}-\chi^2_{\rm opt,\Lambda CDM} \simeq -5.29,
 \label{eq:deltachi-fitcheck}
\end{equation}
where $\chi^2=-2\ln\mathcal{L}$, evaluated with the same likelihood in
both models.

The negative difference denotes a lower evaluated objective for F42 in this analysis. It is not a Bayesian evidence difference. Moreover, the fixed-exponent F42 family does not contain $\Lambda$CDM as a parameter limit, so a nested-model one-degree-of-freedom significance must not be assigned to $|\Delta\chi^2|$ merely because F42 has one more fitted parameter.

This result should not be numerically identified with the $\Delta\chi^2\simeq-13.5$ reported for CMB+DESI DR2 BAO+Union3 in Ref.~\cite{Borghetto:2026x}. The archived calculation here omits Union3 and explicitly fixes the massless neutrino sector. A difference in the data combination or neutrino prescription changes the likelihood surface; these records do not isolate a numerical contribution from either change.
The CMB+BAO and CMB+BAO+Union3 analyses of that reference select nearby F42 regions, $w_0\simeq-0.8$ and $w_0\simeq-0.79$, respectively. The present $w_0 \simeq -0.79$ is a nearby benchmark, but agreement of parameter regions is not itself a quantitative validation of the different $\Delta\chi^2$ values.

\subsection{Linear growth}
\label{sec:growth}

The linear growing mode $D(a)$ relates equal early perturbations to the amplitude before nonlinear mode coupling becomes important. Its logarithmic rate is $f(a) \equiv \dd\ln D/\dd\ln a$. These two quantities connect the clock
to the power response; their relative responses need not have the same sign.

In the late-time single-dust approximation, Eqs.~\eqref{eq:gr-subhorizon-limits} show that both the gravitational source and the friction coefficient reduce to their standard-force forms at leading subhorizon order. 
At this order the growing mode therefore obeys the scale-independent standard-force equation on each model's background:
\begin{equation}
 D''+\left(2+\frac{H'}{H}\right)D'-\frac32\Omega_m(a)D=0 .
 \label{eq:growth}
\end{equation}
We retain radiation in the background functions $H(a)$ and $\Omega_m(a)$, but neglect radiation perturbations in this dust growth equation. We initialize the solution at $z=200$ with a M\'esz\'aros growing-mode slope and normalize the F42 and $\Lambda$CDM amplitudes to agree there. 

\section{Collapse and the large-deviation PDF}
\label{sec:ldt}

We study the dimensionless matter density in spherical cells of comoving radius $R$, $\rho_R(\boldsymbol x,z)\equiv\rho_{m,R}(\boldsymbol x,z)/\bar\rho_m(z)
=1+\delta_R(\boldsymbol x,z)$, where $\boldsymbol x$ is the cell center. At fixed $R$ and redshift $z$, its one-point PDF describes the distribution of $\rho_R$ over cell centers. Starting from Gaussian initial conditions, LDT models this PDF using the linear variance, which determines the statistical weight of initial fluctuations, and spherical collapse, which relates those fluctuations to the final density~\cite{Bernardeau:2016, Uhlemann:2016, Cataneo:2022}. In our implementation, the nonlinear variance of the log-density is supplied by the simulations to set the distribution's width. We describe these ingredients and their combination below.

\subsection{Linear variance and mass conservation}

The variance $\sigma_{\rm L}^2(R,z)\equiv\langle\delta_{\rm L,R}^2\rangle$ of the linear matter contrast smoothed within spheres of comoving radius $R$ is given by
\begin{equation}
 \sigma_{\rm L}^2(R,z)= \int_0^\infty\frac{\dd k}{2\pi^2}\,k^2P_{\rm lin}(k,z)\,W^2(k \, R),
 \label{eq:linear-variance}
\end{equation}
where $P_{\rm lin}$ is the linear matter power spectrum and
\begin{equation}
 W(x)=3\,\frac{\sin x-x\cos x}{x^3}
 \label{eq:window}
\end{equation}
is the Fourier transform of the normalized spherical top-hat window. With massless neutrinos, $P_{\rm lin}$ is the linear power spectrum of the total nonrelativistic matter (CDM and baryons).

Let $\tau$ denote the initial linear density contrast of a spherical
Lagrangian region, extrapolated by linear growth to redshift $z$. For a fixed
Lagrangian smoothing radius $R_{\rm L}$, Gaussian initial conditions imply
a probability density proportional to $\exp(-I_{\rm L})$, where
\begin{equation}
 I_{\rm L}(\tau; R_{\rm L},z) = \frac{\tau^2}{2\sigma_{\rm L}^2(R_{\rm L},z)}.
 \label{eq:gaussian-cost}
\end{equation}
This Gaussian rate function quantifies the statistical cost of a fluctuation
$\tau$ at fixed $R_{\rm L}$ and $z$.

Spherical collapse maps this initial contrast to a final cell density
$\rho$, as described in Sec.~\ref{sec:collapse}. The corresponding
Lagrangian radius $R_{\rm L}$ is defined by the radius enclosing the same
mass at the cosmic mean density. Since matter is conserved and
$\bar\rho_m(a)a^3$ is constant, mass conservation gives
$R_{\rm L}^3=R^3 \, \rho$ for a final Eulerian cell of comoving radius $R$:
\begin{equation}
 \Rlag=R \, \rho^{1/3}.
 \label{eq:lagrangian-radius}
\end{equation}
Overdense cells therefore correspond to larger Lagrangian regions
($\Rlag>R$) and underdense cells to smaller ones. Because this scale depends
on the final density, the PDF probes the shape of the linear power spectrum,
not only its amplitude.

\subsection{Spherical collapse}
\label{sec:collapse}

In the spherical saddle-point approximation of LDT, statistical isotropy
motivates a spherically symmetric collapse trajectory for a prescribed cell
density~\cite{Bernardeau:1994,Valageas:2001zr,Ivanov:2018lcg}. We model the
region as a uniform sphere of pressureless matter. In the weak-field,
deep-subhorizon Newtonian approximation adopted for VCDM, the standard dust
continuity and Euler equations, together with the leading Newtonian Poisson
constraint, yield the nonlinear spherical-collapse equation
\cite{Bernardeau:2002, Schmidt:2008tn}:
\begin{equation}
 \delta'' + \left(2 + \frac{H'}{H} \right) \delta' -\frac{4}{3}\frac{(\delta')^2}{1+\delta}
 -\frac{3}{2} \Omega_m \delta(1 + \delta) = 0.
 \label{eq:collapse}
\end{equation}
Here $\delta$ is the nonlinear top-hat matter contrast, with
$\rho=1+\delta$ at the final epoch. Within this approximation, the
background enters only through $H(a)$ and $\Omega_m(a)$; radiation affects
the expansion but is treated as homogeneous in the collapse calculation.
We solve Eq.~\eqref{eq:collapse} with the background of each model,
Eqs.~\eqref{eq:f42-E} and \eqref{eq:lcdm-E}, using the standard-force
Newtonian closure for the nonlinear matter field.

We start the integration at $a_i=10^{-4}$ from an initial linear growing-mode
amplitude $\epsilon\equiv\delta_{\rm lin}(a_i)$, including the effect of
radiation on the initial growth rate (see Appendix~\ref{app:new-collapse-check}
for the nonlinear initial conditions). For each target density $\rho$ at
$a_f \equiv (1+z)^{-1}$, we adjust $\epsilon$ until the integrated solution obeys
$1+\delta(a_f;\epsilon)=\rho$. The same initial fluctuation, extrapolated
linearly to the final redshift using the growth factor $D(a)$ introduced in
Sec.~\ref{sec:growth}, is
\begin{equation}
 \tau \equiv \epsilon \left[\frac{D(a_f)}{D(a_i)} \right] .
 \label{eq:tau-def}
\end{equation}
This defines the inverse spherical-collapse map $\tau_M(\rho,z)$ for each
suite $M$ ($V$, $\Lambda_{V}$, and $\Lambda_\mathrm{fit}$). We have checked its insensitivity to the choice of $a_i$ and to
the numerical integration accuracy. For comparison (see Fig.~\ref{fig:collapse-map-v9} below), we also use the
approximation of Ref.~\cite{Bernardeau:1994} to the collapse map in an
Einstein--de Sitter (EdS) universe,
\begin{equation}
 \tau_{\rm EdS}(\rho)=\nu\left(1-\rho^{-1/\nu}\right),
 \qquad \nu \equiv \frac{21}{13}.
 \label{eq:eds-map}
\end{equation}

\subsection{Rate function and PDF}
\label{sec:pdf-model}

The preceding ingredients determine the rate function of the final
density. By the contraction principle of LDT, the leading statistical cost
of a final density $\rho$ is inherited from the Gaussian initial contrast
$\tau_M(\rho,z)$ at the associated Lagrangian scale
$R_{\rm L}=R\rho^{1/3}$~\cite{Bernardeau:2016,Uhlemann:2016}.
We define the reduced one-cell rate function for suite $M$ as
\begin{equation}
 \Psi_{M,R}(\rho,z) \equiv \left[ \frac{\sigma_{M,{\rm L}}^2(R,z)}{2\sigma_{M,{\rm L}}^2(R\rho^{1/3},z)} \right] \tau_M^2(\rho,z).
 \label{eq:rate}
\end{equation}
The factor $\sigma_{M,{\rm L}}^2(R,z)$ extracts the overall linear-variance
amplitude from the Gaussian cost. Only the ratio of variances at two scales
enters $\Psi$, so a scale-independent rescaling of the linear power
spectrum cancels from this reduced rate function. The overall width of the
finite-variance PDF is supplied separately below; it is not predicted by
this cancellation.

The PDF follows from the reduced rate function through a saddle-point
approximation. Working with the log-density $\mu \equiv \ln\rho$, for which
the rate function is more nearly convex, and transforming the result back
to $\rho$ gives~\cite{Uhlemann:2016,Cataneo:2022}
\begin{equation}
 \widetilde\Pc_{M,R}(\rho)=
 \sqrt{\frac{\Psi_{M,R}''(\rho)+\Psi_{M,R}'(\rho)/\rho}
 {2\pi s_{M,R}^2}}
 \exp\!\left[-\frac{\Psi_{M,R}(\rho)}{s_{M,R}^2}\right] ,
 \label{eq:raw-pdf}
\end{equation}
where primes in this equation denote derivatives with respect to $\rho$,
unlike the time derivatives used earlier. The positive parameter
$s_{M,R}^2$ controls the width of the finite-variance PDF. It is neither
automatically equal to $\sigma_{M,{\rm L}}^2(R,z)$ nor simply set equal to
the measured log-density variance before normalization; instead, it is
calibrated by matching the measured log-density variance as described below.

The saddle-point formula is used only where the rate function in $\mu$ has positive curvature. Since it does not exactly integrate to one or have unit mean, we normalize it and rescale the density following Refs.~\cite{Uhlemann:2017,Cataneo:2022}. At fixed model, radius, redshift and trial width $s_{M,R}^2$, we define \[
\begin{gathered}
 I_{n,M,R}=\int_0^\infty\dd r\,r^n\widetilde\Pc_{M,R}(r),
 \qquad n=0,1,\\
 m_{M,R}=\frac{I_{1,M,R}}{I_{0,M,R}}.
\end{gathered}
\]
The corrected PDF is
\begin{equation}
 \Pc_{M,R}(\rho)=
 \frac{m_{M,R}}{I_{0,M,R}}\,
 \widetilde\Pc_{M,R}(m_{M,R}\rho),
 \label{eq:tilted-pdf}
\end{equation}
which satisfies
\begin{equation}
 \int\dd\rho\,\Pc_{M,R}(\rho)=1,
 \qquad
 \int\dd\rho\,\rho\,\Pc_{M,R}(\rho)=1.
 \label{eq:unit-mean}
\end{equation}
Equivalently, the normalized log-density distribution is shifted by $-\ln m_{M,R}$, which preserves its variance and higher cumulants. 

Finally, for each model, redshift and radius, we determine $s_{M,R}^2$ by matching the measured, CIC-deconvolved log-density variance, averaged over the realizations,
\begin{equation}
 \Var_{\Pc_{M,R}}(\ln\rho)=\widehat{\Var}_M(\ln\rho_R),
 \label{eq:width-match}
\end{equation}
without shot-noise subtraction or fitting the PDF shape. Since the shift preserves the variance, the width can be calibrated before the density rescaling is applied. LDT therefore predicts the shape of the PDF response for the measured width, not the change in the width itself.

\section{Linear input, matched simulations, and estimators}
\label{sec:numerics}

In this section, we implement the preceding formalism via a three-stage numerical pipeline: first, the generation of linear matter power spectra with \class; second, the execution of $N$-body simulations using \mgcola; and third, the measurement of smoothed fields and response estimators. Table~\ref{tab:simulation} summarizes the simulation configuration.

\begin{table}[t]
 \caption{Simulation and analysis settings. The $\Lambda_{\rm fit}$ box  and radii are rescaled by $\alpha_h\equiv h_{\Lambda_{\rm fit}}/h_V$ so that all suites have the same physical box size and smoothing radii.}
 \label{tab:simulation}
 \centering
 \begin{ruledtabular}
 \begin{tabular}{ll}
 Quantity & Value \\
 \hline
 Realizations per suite & $20$ \\
 Particles and force-mesh cells & $1024^3$ each \\
 $V$ and $\Lambda_V$ box side & $2\,h_V^{-1}\,\mathrm{Gpc}$ \\
 $\Lambda_{\rm fit}$ box side & $\approx2.063\,h_{\Lambda_{\rm fit}}^{-1}\,\mathrm{Gpc}$ \\
 Initial redshift & $z_\mathrm{ini}=19$ \\
 Timesteps & $40=19+5+2+14$ \\
 Output redshifts & $1$, $z_\times=0.61$, $0.5$, $0$ \\
 $V$ and $\Lambda_V$ smoothing radii & $10$, $15$, $20$, $25\,h_V^{-1}\,\mathrm{Mpc}$ \\
 $\Lambda_{\rm fit}$ smoothing radii & $\alpha_h R_V$, with $\alpha_h\approx1.0316$ \\
 Analysis $k$ range & $k\in[0.02,0.2]\,h_V\,\mathrm{Mpc}^{-1}$ \\
 Paired bootstrap resamples & $10^4$ \\
 \end{tabular}
 \end{ruledtabular}
\end{table}

\subsection{Linear power spectra and simulation resolution}

We compute linear matter power spectra with the Boltzmann solver \class{} v3.2.1~\cite{Blas:2011}\footnote{\url{http://class-code.net/}}, in which we implemented the F42 background of VCDM. For the $V$ suite, the sampled acoustic-scale parameter $100\theta_s \approx 1.04$ yields $H_0\approx66.57\,\mathrm{km\,s^{-1}\,Mpc^{-1}}$ in \class{} (Table~\ref{tab:cosmology}). This derived $H_0$ is supplied directly to the \LCDM{} control $\Lambda_{V}$, rather than imposing the same $100\theta_s$ in both backgrounds. At each required output redshift, we use the corresponding linear spectrum, tabulated on a 4096-point grid spanning $1.06\times10^{-5}\le k/(\kMpc)\le167.53$ in each model's $h$-scaled units, with interpolation in $\ln k$--$\ln P$ and no extrapolation. These spectra provide the linear inputs for the \mgcola{} simulations.

For the $V$ and $\Lambda_{V}$ suites, with periodic box side $L=2\,h_V^{-1}\mathrm{Gpc}$ and mesh size $N_\mathrm{mesh}=1024$, the fundamental wavenumber $k_\mathrm{f}$ and the mesh Nyquist wavenumber $k_\mathrm{Ny}$ are
\begin{align}
\label{eq:kf-kNy}
    k_\mathrm{f} &\equiv \frac{2\pi}{L} \approx 0.0031\,h_V\,\mathrm{Mpc}^{-1}, \\
    k_\mathrm{Ny} &\equiv \frac{\pi N_\mathrm{mesh}}{L} \approx 1.61\,h_V\,\mathrm{Mpc}^{-1}.
\end{align}
Our primary interval $0.02\le k/(h_V\,\mathrm{Mpc}^{-1})\le0.2$ therefore lies between $6.37\,k_\mathrm{f}$ and $0.12\,k_\mathrm{Ny}$. This separation limits exposure to box-size and grid-sampling effects, but does not by itself establish negligible aliasing or sufficient force resolution~\cite{Sefusatti:2016}.

\subsection{Phase-matched \mgcola{} ensembles on a common comoving volume}
\label{sec:cola}

We perform collisionless $N$-body simulations using \mgcola{}~\cite{Tassev:2013, Howlett:2015hfa, Winther:2017, Wright:2017dkw} with standard Newtonian particle forces and the background expansion and growth functions appropriate to F42 or \LCDM. The simulations are organized into three suites: $V$, $\Lambda_{V}$, and $\Lambda_{\rm fit}$. Suites $V$ and $\Lambda_{V}$ share the dimensionless Hubble parameter $h_V$, whereas suite $\Lambda_{\rm fit}$ uses $h_{\Lambda_{\rm fit}} \neq h_V$. To preserve an identical physical comoving box side $L_\mathrm{com} = b_V h_V^{-1} \, \mathrm{Mpc} \approx 3004.5 \, \mathrm{Mpc}$ across all runs, we scale the code-unit box length as $b_{\Lambda_{\rm fit}} = \alpha_h b_V \approx 2063.26$, where $\alpha_h \equiv h_{\Lambda_{\rm fit}}/h_V \approx 1.03$ and $b_V = b_{\Lambda_V} = 2000$. The common box length in comoving Mpc gives the same physical Fourier-mode lattice in all three suites. Using the same mode-indexed random realization then matches the initial phases, while the mode amplitudes retain the model-dependent power spectra.

Each suite comprises 20 realizations with matched random seeds (5005--5024). Every run evolves $1024^3$ particles on a $1024^3$ force mesh. Starting at $z_\mathrm{ini}=19$, the production simulations use 40 COLA timesteps distributed as $19+5+2+14$ between successive outputs at $z=1$, $z_\times$, $z=0.5$, and $z=0$, respectively. The archived production schedule specifies approximately uniform spacing in the scale factor $a$, while imposing the four output redshifts, including the phantom crossing, as interval boundaries. \mgcola{} advances with constant $\Delta a$ within each interval, with small changes in the step size between intervals to reach the prescribed endpoints. All three suites use this same schedule. A dedicated one-seed check of the $z=0$ power spectrum response against finer timestep counts is reported in Appendix~\ref{app:validation}; it does not establish convergence of the full PDF or every epoch.

To compare models with different $h$ at common comoving scales, we convert power spectra and wavenumbers from the conventional $h$-scaled units to Mpc units:
\begin{equation}
    P_{\mathrm{phys}} = P_h/h^3, \qquad k_{\mathrm{phys}} = h k_h.
\label{eq:physical-pk-conversion}
\end{equation}
Similarly, spatial smoothing must be applied at identical physical
comoving scales. Taking $R_V\in\{10,15,20,25\}\,h_V^{-1}\,\mathrm{Mpc}$ as the baseline for the $V$ and $\Lambda_V$ suites, we scale
$R_{\Lambda_{\rm fit}}=\alpha_h R_V$ so that all suites are smoothed over identical physical radii $R\approx\{15.02,22.53,30.04,37.56\}\,\mathrm{Mpc}$. Here ``physical'' refers to comoving Mpc units, not proper lengths. At the smallest radius, the mesh cell is about $R/5$; a resolution test with cells of about $R/10$ is described in Appendix~\ref{app:mesh-ref}. For presentation, measured spectra are interpolated onto a finer common grid; this interpolation does not create additional independent Fourier modes.
\begin{figure}[t]
 \centering
 \includegraphics[width=\columnwidth]{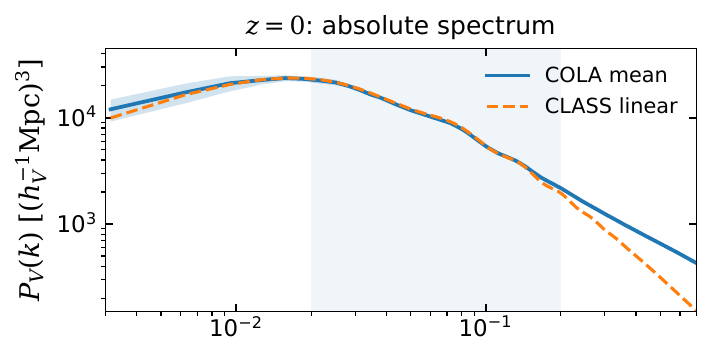}\\[2pt]
 \includegraphics[width=\columnwidth]{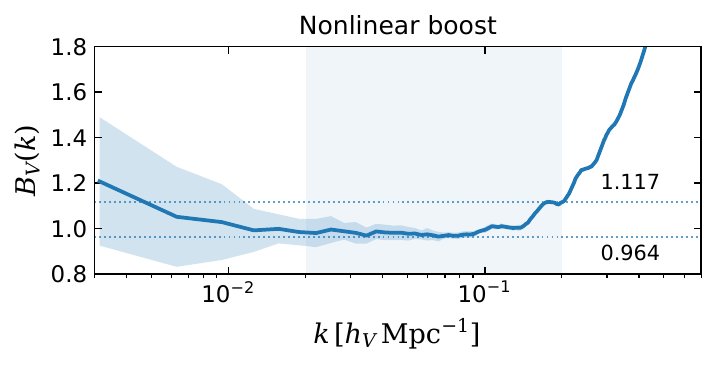}\\[2pt]
 \caption{Absolute power spectrum consistency diagnostic at $z=0$. \textbf{Top:} $V$-suite ensemble mean and sample standard deviation across the 20 realizations, compared with its \class{} linear spectrum. \textbf{Bottom:} Nonlinear boost $B_V(k) = P_{V,\mathrm{COLA}}(k)/P_{V, \mathrm{lin}}(k)$; the two explicitly labeled dotted guides give approximately its primary-range extrema $0.96$ and $1.12$. The shaded band marks $0.02\leq k/(h_V\,\mathrm{Mpc}^{-1})\leq0.2$.
 }
 \label{fig:absolute-power}
\end{figure}

To validate our numerical setup, we inspect the $z=0$ absolute power spectrum of the $V$ ensemble, $P_{V, \mathrm{COLA}}(k)$, against linear theory predictions from \class{}, $P_{V, \mathrm{lin}}$ (see Fig.~\ref{fig:absolute-power}). On large scales, the ensemble mean tracks linear theory within realization scatter. At higher $k$, the spectrum displays the expected nonlinear upturn driven by mode coupling (top panel). Across the analysis $k$-range (shaded region), the nonlinear boost, $B_V(k) \equiv P_{V, \mathrm{COLA}} / P_{V, \mathrm{lin}}$, spans $0.96$--$1.12$ (labeled horizontal guides in the bottom panel). This is consistent with the expected linear behavior on large scales and with nonlinear evolution at smaller scales; by itself, it is not an absolute force-resolution or nonlinear-accuracy test. The increased scatter at the lowest modes is consistent with the limited number of Fourier modes in the finite simulation volume.

\subsection{Summary statistics and estimators}
\label{sec:estimators}

\begin{figure*}[t]
 \centering
 \includegraphics[width=0.318\textwidth]{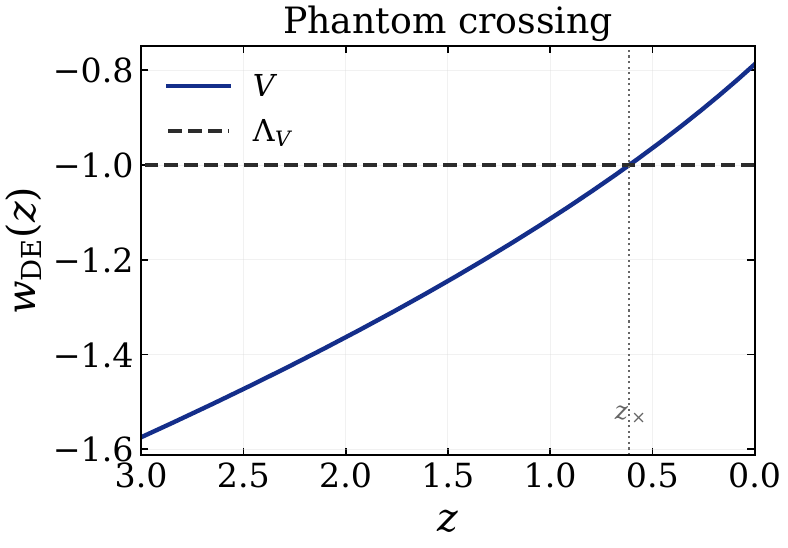}
\hfill
 \includegraphics[width=0.318\textwidth]{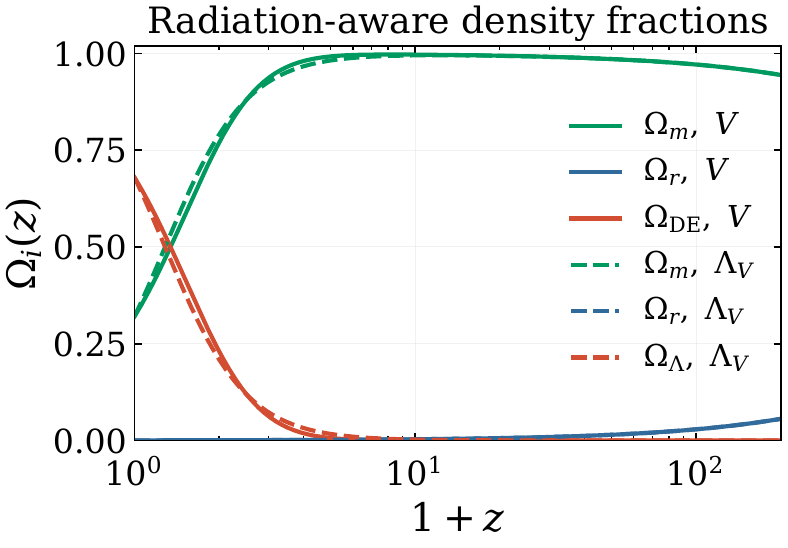}
\hfill
 \includegraphics[width=0.318\textwidth]{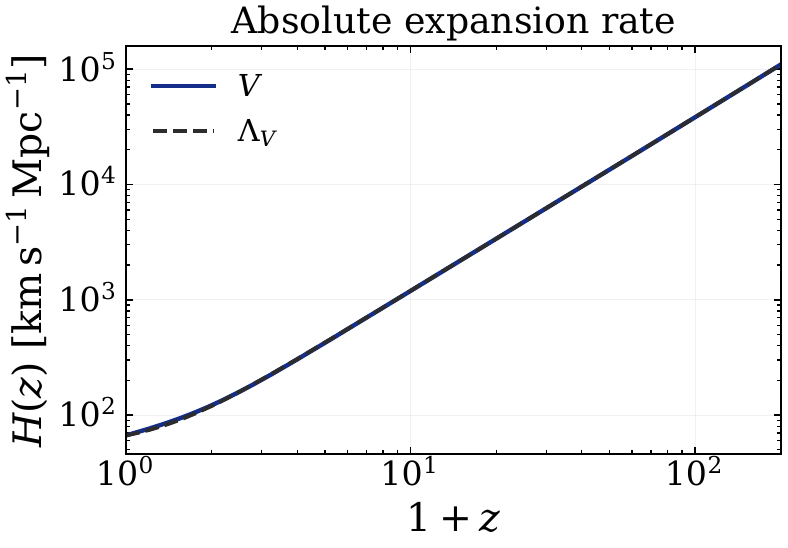}
\\[4pt]
 \includegraphics[width=0.318\textwidth]{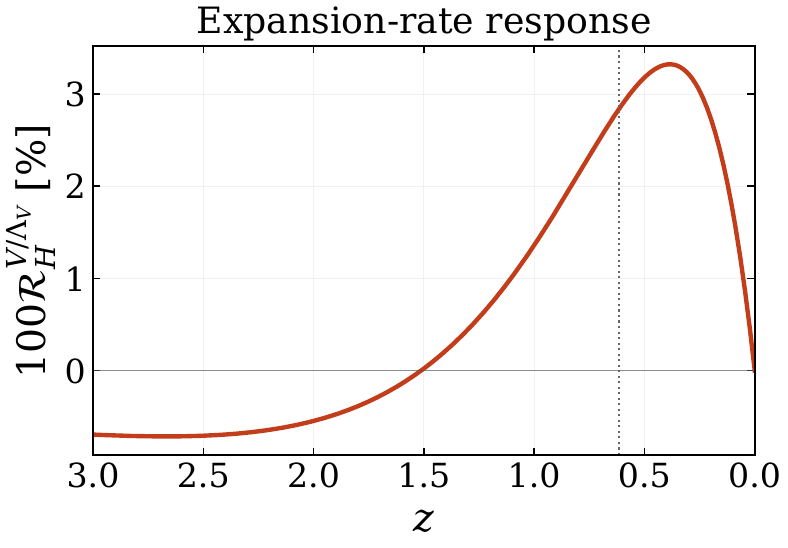}
\hfill
 \includegraphics[width=0.318\textwidth]{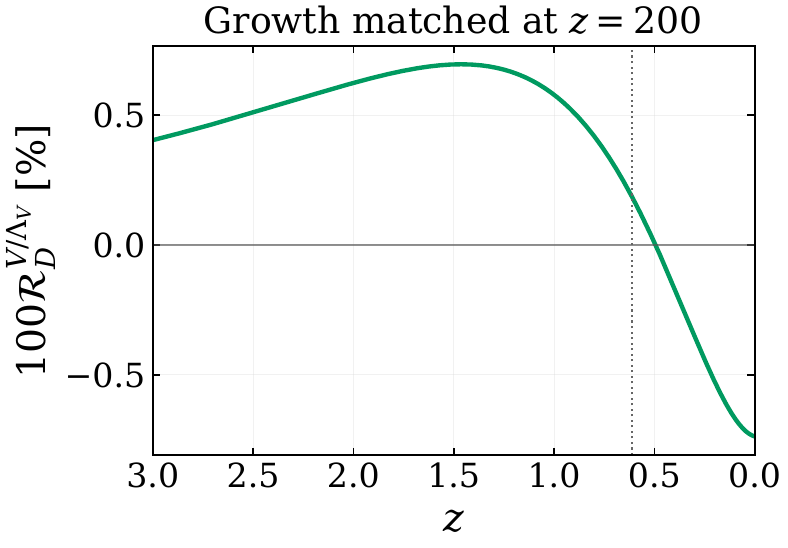}
\hfill
 \includegraphics[width=0.318\textwidth]{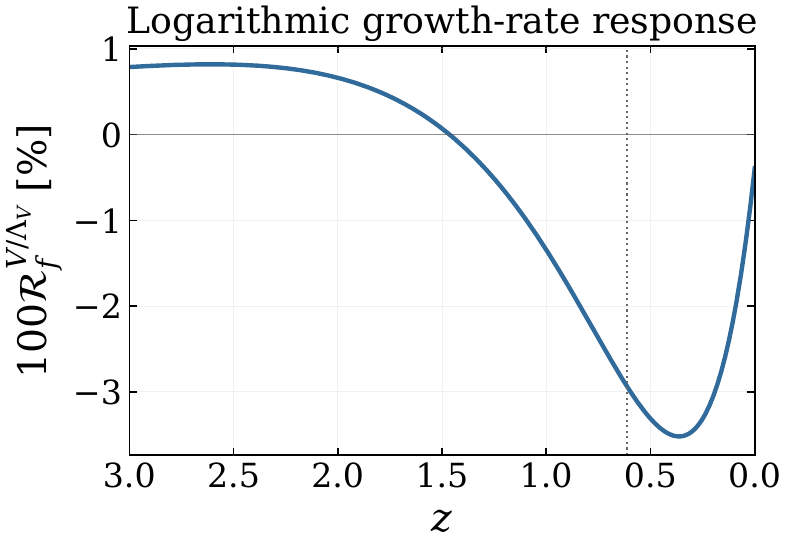}
 \caption{Background evolution (top row) and responses (bottom row) of
 $V$ (F42) relative to $\Lambda_V$ (\LCDM{}). \textbf{Top:} effective
 equation of state $w_\phi$, density fractions of matter (green),
 radiation (blue) and dark energy (red), and Hubble rate $H(z)$; solid
 curves show $V$ and dashed curves $\Lambda_V$. \textbf{Bottom:}
 responses of the Hubble rate, $100\Resp_H^{V/\Lambda_V}$, the growth
 factor, $100\Resp_D^{V/\Lambda_V}$, and the growth rate,
 $100\Resp_f^{V/\Lambda_V}$, with growth amplitudes matched at $z=200$.
 Vertical dotted lines mark the phantom crossing. In the panels plotted
 against $z$, cosmic time increases from left to right; the density
 fractions and $H(z)$ are plotted against $1+z$ on a logarithmic axis.
 }
 \label{fig:background-growth}
\end{figure*}

To quantify the nonlinear matter distribution, we measure the power spectrum $P(k)$, three cell variances (the Fourier-space variance, the real-space density variance and the log-density variance), and the one-point PDF of $\mu_R\equiv\ln\rho_R$. Particle masses are assigned to a $1024^3$ mesh with the cloud-in-cell (CIC) scheme and then smoothed with a spherical top-hat. Our main estimators correct for the CIC assignment window (deconvolution), while the uncorrected fields are used as a check. Poisson shot noise is subtracted from the real-space density variance, while the log-density variance used to calibrate the LDT PDF prediction is not corrected for shot noise, since in the latter case discreteness changes the shape of the distribution and cannot be removed by a simple subtraction. Numerical checks of the estimators and of the analysis are summarized in Appendix~\ref{app:validation}.

All statistics are measured at $z=1$, $z_\times=0.613$, $0.5$ and $0$. The power spectrum is measured in 57 bins over $0.02\leq k/(h_V\,\mathrm{Mpc}^{-1})\leq0.2$, approximately $0.013\leq k/\mathrm{Mpc}^{-1}\leq0.13$ in physical units. The cell variances and the PDF are measured in cells of radius $R\approx15$, $23$, $30$ and $38\,{\rm Mpc}$. For the PDF, we use log-density bins in the range $-3\leq\mu\leq3$. 

To compare a suite $A$ with a suite $B$, we evaluate for each of the $N_{\mathrm{pair}}=20$ phase-matched realization pairs $s$ the per-pair log-deviation
\begin{equation}
 \Dev^{A/B}_{X,s}(q) \equiv \ln\left[\frac{X_{A,s}(q)}{X_{B,s}(q)}\right],
 \label{eq:paired-deviation}
\end{equation}
where $X$ represents the power spectrum $P(k)$, the nonlinear boost $B(k)\equiv P_{\mathrm{COLA}}/P_{\mathrm{lin}}$, the cell variances $\sigma^2$, or the log-density PDF  $\mathcal{P}_\mu{(\mu)}$ and $q$ indexes the variable associated to $X$. Log ratios convert multiplicative effects into additive deviations, while matched phases reduce the realization scatter, especially between $V$ and $\Lambda_V$. The ensemble mean log-deviation is the logarithm of the geometric mean of the per-seed ratios, not generally the logarithm of a ratio of
arithmetic ensemble means,
\begin{equation}
 \mDev^{A/B}_X(q) \equiv \frac{1}{N_\mathrm{pair}}
   \sum_{s=1}^{N_\mathrm{pair}} \Dev^{A/B}_{X,s}(q).
\end{equation}
Its standard error at each coordinate $q$ is obtained from the sample variance of the paired log-deviations,
\begin{equation}
 \mathrm{SE}\big[\mDev^{A/B}_X(q)\big] =
 \left\{\frac{\sum_{s=1}^{N_{\mathrm{pair}}}
 \left[\Dev^{A/B}_{X,s}-\mDev^{A/B}_X\right]^2}
 {N_{\mathrm{pair}}(N_{\mathrm{pair}}-1)}\right\}^{1/2}.
 \label{eq:response-estimator}
\end{equation}
Finally, we convert back to a fractional change, or response,
\begin{equation}
 \Resp^{A/B}_X(q) \equiv \exp\!\left[\mDev^{A/B}_X(q)\right] - 1.
 \label{eq:response-notation}
\end{equation}
In the results, power-spectrum panels show $100\Resp_X$, whereas the PDF panels show the logarithmic response $100\mDev_{\mathcal P}$; these
coincide only to first order in a small response.

The bands in the figures show $95\%$ intervals for the mean log responses, computed separately for each coordinate from $10^4$ paired bootstrap resamples of the 20 independent seeds. We resample seeds rather than cells, because cells from the same simulation overlap strongly and are not independent. These intervals include only the scatter between realizations, not systematic effects such as resolution
or baryonic physics.

We use the three suites to separate two effects. The clock response, comparing $V$ with $\Lambda_V$, isolates the difference between the F42 and \LCDM{} expansion histories at fixed parameters, while the refit response, comparing $\Lambda_V$ with $\Lambda_{\rm fit}$, captures the shift of the $\Lambda$CDM parameters. Because log responses add, their sum gives the total difference between the two fitted models,
\begin{equation}
 \underbrace{\mDev_X^{V/\Lambda_{\rm fit}}}_{\text{total}}
 =\underbrace{\mDev_X^{V/\Lambda_V}}_{\text{clock}}
 +\underbrace{\mDev_X^{\Lambda_V/\Lambda_{\rm fit}}}_{\text{refit}}.
 \label{eq:clock-refit-log}
\end{equation}

\section{Results}
\label{sec:results}

We first examine how the F42 expansion history affects the linear growth of structure, comparing $V$ with $\Lambda_V$. We then present the measured response of the power spectrum, and use the three suites $V$, $\Lambda_V$ and $\Lambda_{\rm fit}$ to separate the effect of the expansion history (the clock) from that of refitting the parameters. Before turning to the PDF, we show that the spherical-collapse map is nearly unaffected by the change of background. We then present the responses of the cell variances and the one-point PDF, and finally compare them with the simulation-calibrated LDT prediction.

\subsection{The background clock and its growth memory}
\label{sec:background-results}

\begin{table}[t]
 \caption{Hubble parameter, growth factor, and growth rate responses of $V$ relative to its same-parameter \LCDM{} control $\Lambda_{V}$ at the four simulation epochs. The growth factors are normalized to equal amplitude at $z=200$.}
 \label{tab:clock}
 \centering
 \begin{ruledtabular}
 \begin{tabular}{lrrr}
 Epoch & $100\Resp_H$ & $100\Resp_D$ & $100\Resp_f$ \\
 \hline
 $z=1$ & $+1.360$ & $+0.578$ & $-1.344$ \\
 $z=z_\times$ & $+2.846$ & $+0.188$ & $-2.933$ \\
 $z=0.5$ & $+3.180$ & $+0.009$ & $-3.311$ \\
 $z=0$ & $0.000$ & $-0.738$ & $-0.393$ \\
 \end{tabular}
 \end{ruledtabular}
\end{table}

Figure~\ref{fig:background-growth} compares the effective equation of state, density fractions, expansion rate and linear-growth responses of $V$ (F42) and its matched \LCDM{} control $\Lambda_V$. In the first panel of the top row, $w_\phi$ rises with cosmic time and crosses $-1$ at $z_\times$, whereas the cosmological constant has $w=-1$ throughout. The second panel follows the density fractions to $z=200$. The two models
have the same fractions today but different histories at intermediate redshifts, where the effective dark-energy fraction is larger in $V$ and $\Omega_m$ correspondingly lower, by up to about $6\%$ near $z\simeq0.5$.
At $z=200$, radiation still contributes about $5.60\%$ of the energy budget in both models, which is why we include it in the background and in the initialization of the collapse calculation. The third panel shows $H(z)$.

The bottom panels show the background and growth responses. These are deterministic, so no seed averaging is needed. The Hubble-rate response $100\Resp_H$ is slightly negative at early times, crosses zero near $z\approx1.51$, and reaches a maximum of $+3.32\%$ at $z\approx0.38$. It then returns to zero today, since $V$ and $\Lambda_V$ share the same $H_0$. This maximum occurs after the phantom crossing, and also differs
from the redshift at which the growth response vanishes.

The other two panels show the responses of the linear growth factor $D$ and of its logarithmic rate $f$. At fixed physical matter density, a change in $H(a)$ alters both the Hubble friction and the matter fraction $\Omega_m(a)$, and therefore the growth history. With the two growth amplitudes matched at $z=200$, this convention isolates how the two expansion histories evolve the same early perturbation. The accumulated response follows from
\begin{equation}
 \ln\!\left(\frac{D_V(a)}{D_{\Lambda_V}(a)}\right)
 =\int_{a_{200}}^a\!\left(f_V-f_{\Lambda_V}\right)\dd\ln a,
 \label{eq:growth-memory}
\end{equation}
where the ratio is unity at $z=200$. The slower early expansion of $V$ first lets it build a small growth advantage over $\Lambda_V$, which peaks near $z\simeq1.5$. Once $\Resp_H$ becomes positive, growth in $V$
slows, and $\Resp_f$ is most negative between the phantom crossing and today. The advantage therefore shrinks, vanishes near $z\simeq0.5$, and turns into a deficit of $0.74\%$ today. The equal present-day Hubble
rates do not erase this difference, because
Eq.~\eqref{eq:growth-memory} integrates the whole growth-rate history.
In particular, the phantom crossing does not by itself determine the redshift at which the accumulated response vanishes. Representative
values are given in Table~\ref{tab:clock}.

\begin{table*}[t]
\caption{Responses of $V$ relative to $\Lambda_V$ at the four epochs, in
percent. For the power spectrum, the interval gives the minimum and
maximum of the response over $0.02\leq k/(h_V\,\mathrm{Mpc}^{-1})\leq0.2$.
All other columns are given at $R=10\,h_V^{-1}\,\mathrm{Mpc}$. The PDF
column gives the largest measured fractional response, $100\max|\Resp_{\Pc}|$, from the paired log-ratio estimator, over the retained bins within
the central $98\%$ of the distribution. $E_{\rm rms}$ and $A_{\rm rms}$
are the probability-weighted rms LDT residual and amplitude of the
measured log response, defined in Eq.~\eqref{eq:rms-definition}.}
 \label{tab:response}
 \centering
 \begin{ruledtabular}
 \begin{tabular}{lcccccc}
 Epoch & $100\Resp_P$ & $100\Resp_{\Var(\rho)}$ & $100\Resp_{\Var(\mu)}$
        & $100\max|\Resp_{\Pc}|$ & $100E_{\rm rms}$ & $100E_{\rm rms}/A_{\rm rms}$ \\
 \hline
 $z=1$                & $[1.148, 1.230]$   & $1.226$  & $1.059$  & $3.19$ & $0.022$  & $3.37$ \\
 $z_\times\simeq0.61$ & $[0.369, 0.395]$   & $0.395$  & $0.333$  & $1.04$ & $0.009$  & $4.23$ \\
 $z=0.5$              & $[0.010, 0.019]$   & $0.012$  & $0.019$  & $0.09$ & $0.0009$ & $7.18$ \\
 $z=0$                & $[-1.606, -1.429]$ & $-1.577$ & $-1.168$ & $3.83$ & $0.051$  & $6.15$ \\
 \end{tabular}
 \end{ruledtabular}
\end{table*}

\subsection{Power spectrum response}
\label{sec:Power-response}

\begin{figure}[t]
 \centering
 \includegraphics[width=\columnwidth]{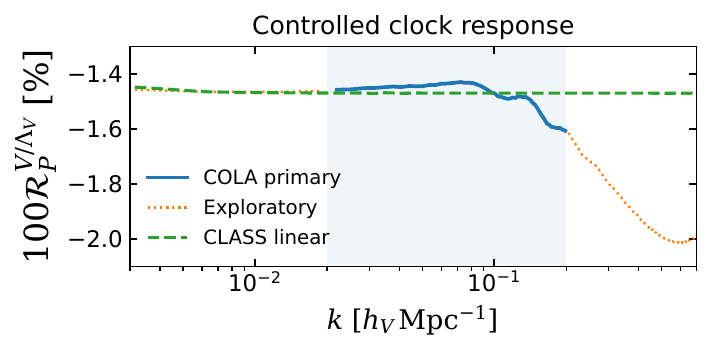}
 \caption{Clock response of the matter power spectrum ($V$ relative to  $\Lambda_V$) at $z=0$, together with its linear prediction. The shaded  band marks $0.02\leq k/(h_V\,\mathrm{Mpc}^{-1})\leq0.2$. The solid blue  line shows the primary response and the dotted orange line its  exploratory extension to smaller scales. Agreement with the linear prediction (dashed green line) is a consistency check.}
 \label{fig:absolute-clock}
\end{figure}

\begin{figure*}[t]
 \centering
 \includegraphics[width=0.485\textwidth]{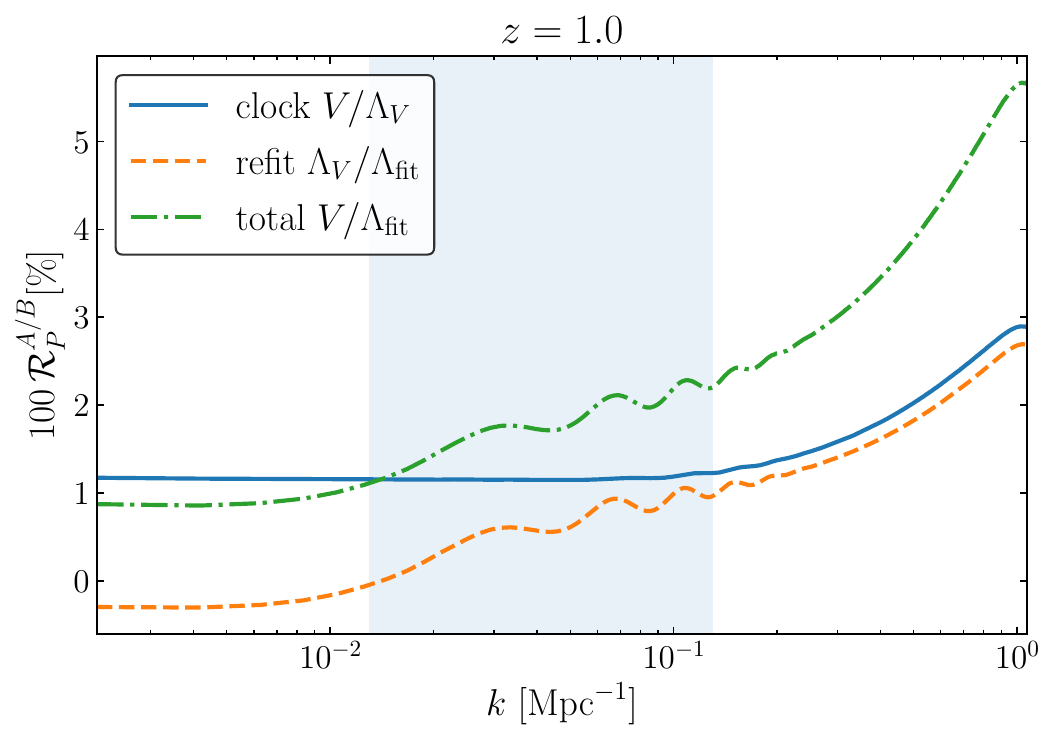}
\hfill
 \includegraphics[width=0.485\textwidth]{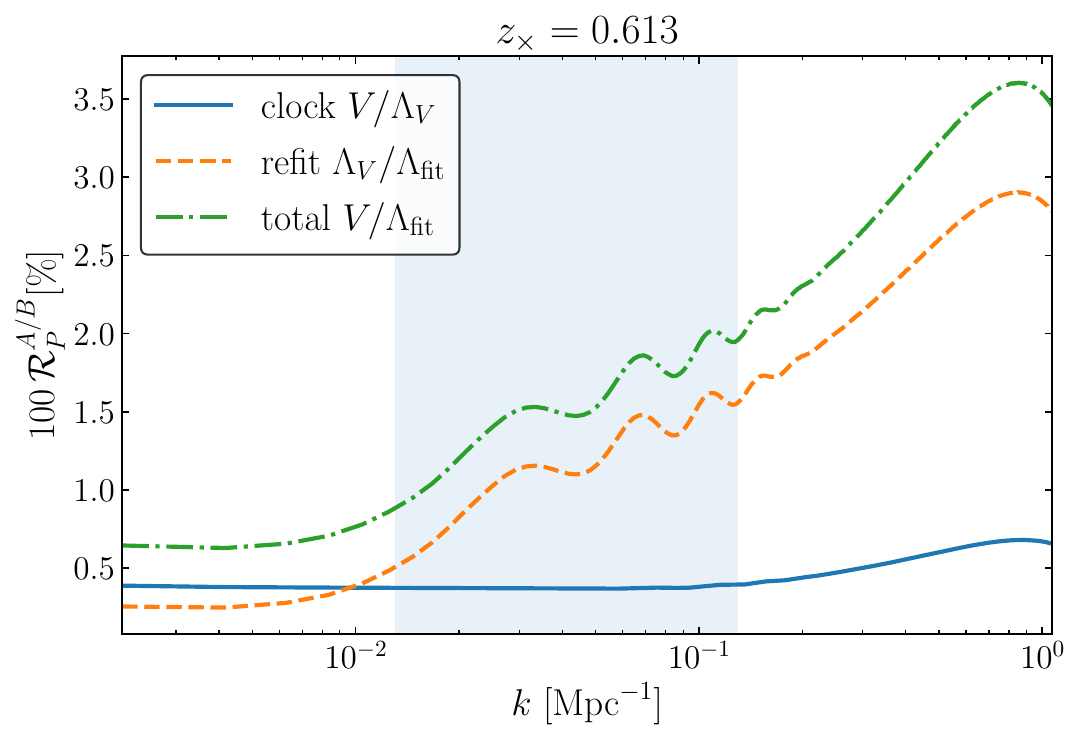}
\\[4pt]
 \includegraphics[width=0.485\textwidth]{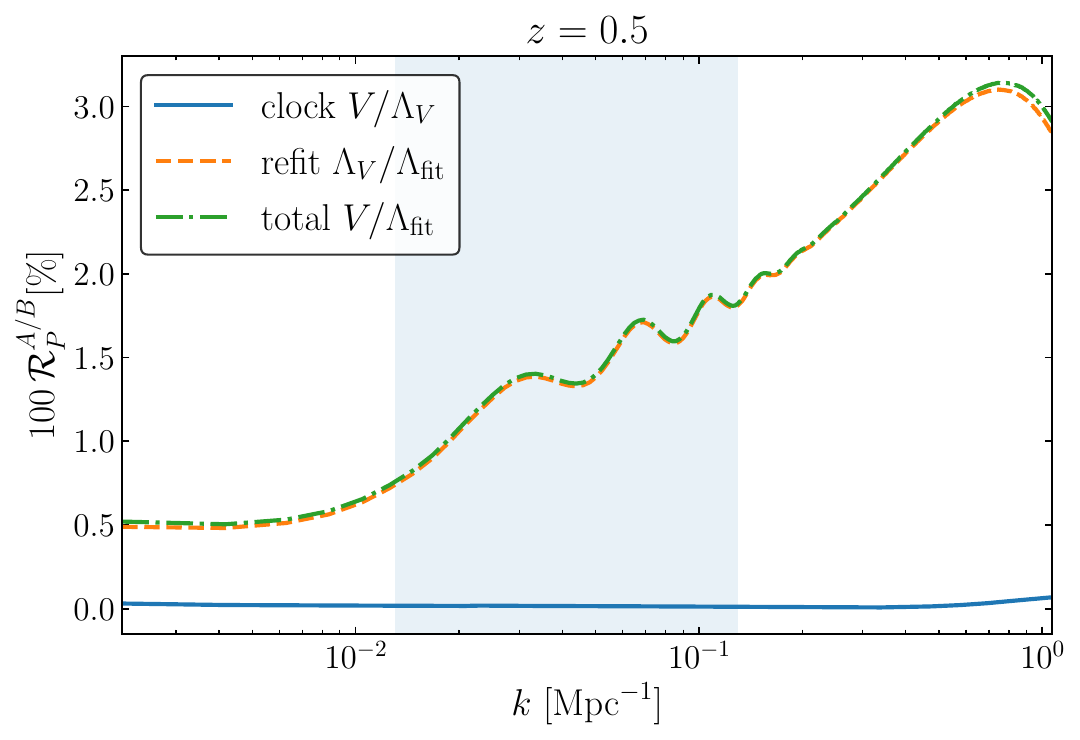}
\hfill
 \includegraphics[width=0.485\textwidth]{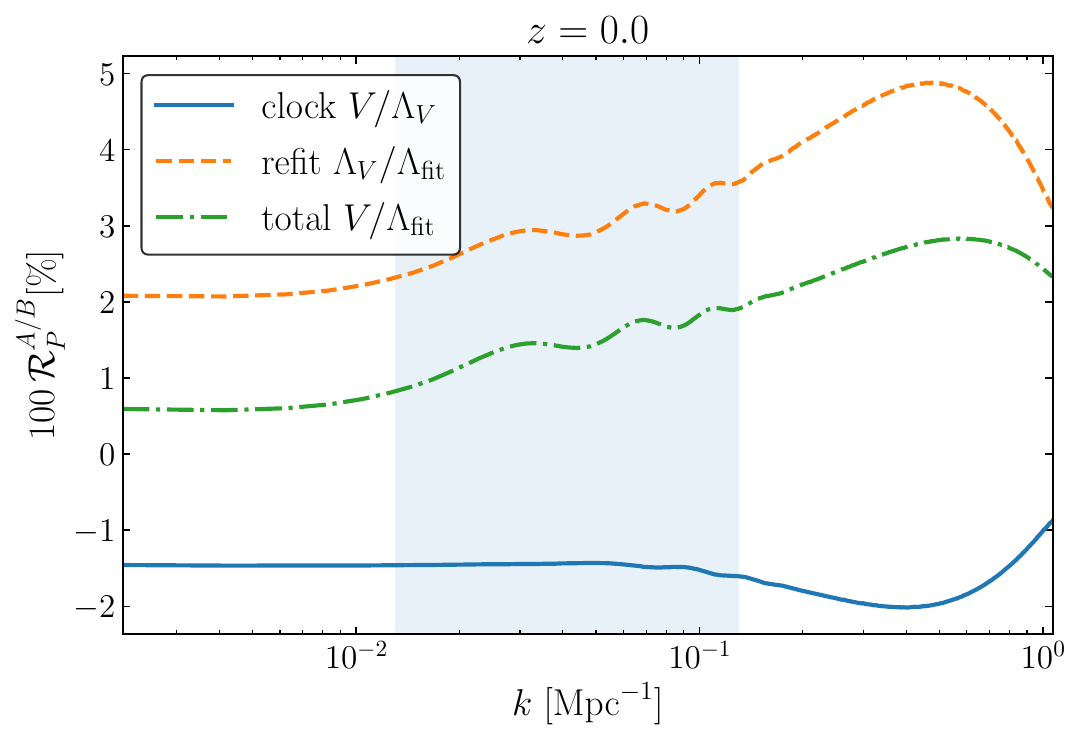}
 \caption{Power spectrum responses at $z=1$, $z_\times$, $z=0.5$ and  $z=0$ (left to right, top to bottom). Blue solid, orange dashed and  green dash-dotted curves show the clock ($V$ relative to $\Lambda_V$),  refit ($\Lambda_V$ relative to $\Lambda_{\rm fit}$) and total ($V$  relative to $\Lambda_{\rm fit}$) responses, respectively. Box rescaling  gives the same Fourier-mode lattice for all suites. The shaded band marks the primary analysis interval; the curves outside it are exploratory.}
 \label{fig:triplet-pk-v9e}
\end{figure*}

We next measure the response of the matter power spectrum from the $N_{\rm pair}=20$ matched realization pairs at the four output epochs. Since growth is approximately scale independent on the scales we analyse, the common early linear spectrum implies $P_{\rm lin}\propto D^2(z)$. For the clock comparison, the linear
prediction is therefore $\Resp_{P,\rm lin}^{V/\Lambda_V}= (1+\Resp_D^{V/\Lambda_V})^2-1\simeq2\Resp_D^{V/\Lambda_V}$.
Over $0.02\leq k/(h_V\,\mathrm{Mpc}^{-1})\leq0.2$, the measured response follows this linear expectation (Table~\ref{tab:response}, compared with Table~\ref{tab:clock}): it is positive at $z=1$, smaller at $z_\times= 0.61$, nearly zero at $z=0.5$, and negative today.

Figure~\ref{fig:absolute-clock} compares the clock response of the power spectrum at $z=0$ with its linear prediction. As expected from the growth response in Fig.~\ref{fig:background-growth} and Table~\ref{tab:clock}, the response is approximately $-1.5\%$ on the large and mildly nonlinear scales analysed. The nonlinear measurement (blue) closely follows the linear prediction (green dashed), showing that the sign-changing signal is mainly inherited from the accumulated linear growth. The extension to smaller scales beyond the analysis range (dotted orange) serves as a cautionary diagnostic rather than a physical claim. Its pronounced scale dependence and turnaround may reflect real nonlinear dynamics, but these scales overlap with those where the force resolution, mesh accuracy and numerical convergence of \mgcola{} require further validation.

The response of the nonlinear boost, $B=P_{\rm COLA}/P_{\rm lin}$, confirms that nonlinear evolution adds only a small correction over the primary $k$ range: $|\Resp_B|$ remains below about $0.07\%$ at $z\geq0.5$ and about $0.14\%$ today. The boost response can have either sign and is much smaller than the percent-level power response at $z=1$ and $z=0$.

Figure~\ref{fig:triplet-pk-v9e} compares the clock, refit and total power responses in common comoving units. The shaded band marks the primary interval, approximately $0.013\leq k_{\rm phys}/\mathrm{Mpc}^{-1}\leq0.13$. The refit response is positive across most of this range, with a small negative part at $z=1$. The total response remains positive at all four epochs, between about $+0.8\%$ and $+2.3\%$ within the analysed interval, while the clock contribution changes sign.  Equation~\eqref{eq:clock-refit-log} gives an exact sum of log responses; for fractional responses, the corresponding relation is $1+\Resp_P^{V/\Lambda_{\rm fit}}=(1+\Resp_P^{V/\Lambda_V})(1+\Resp_P^{\Lambda_V/\Lambda_{\rm fit}})$. Consequently, the total response lies above the refit response at $z=1$ and at the crossing, nearly coincides with it at $z=0.5$, and lies below it today. When the two fitted models are compared, the parameter shift therefore dominates the total power response, with a smaller but clearly sign-changing clock contribution.

The refit response rises broadly with $k$, although acoustic features and small local dips prevent a strictly monotonic trend. Its linear and nonlinear contributions can be separated exactly at the level of log responses,
\begin{equation}
 \mDev_P^{\Lambda_V/\Lambda_{\rm fit}}(k,z)
 =\ln\!\left[\frac{P_{{\rm lin},\Lambda_V}(k,z)}{P_{{\rm lin},\Lambda_{\rm fit}}(k,z)}\right]
 +\mDev_B^{\Lambda_V/\Lambda_{\rm fit}}(k,z).
 \label{eq:linear-boost-split}
\end{equation}
At matched comoving wavenumbers, the linear \class{} ratio accounts for about $82$--$99\%$ of the net rise of the refit log response across the primary interval at the four epochs. The nonlinear boost supplies the rest, although it need not be subdominant at every wavenumber or in every local slope. This predominantly linear trend reflects the \textit{combined} parameter refit, not the contribution of any single parameter~\cite{Eisenstein:1998, Orjuela-Quintana:2024hha}.

\subsection{Spherical collapse is a small shape correction}
\label{sec:collapse-results}

\begin{figure*}[t]
 \centering
 \includegraphics[width=0.485\textwidth]{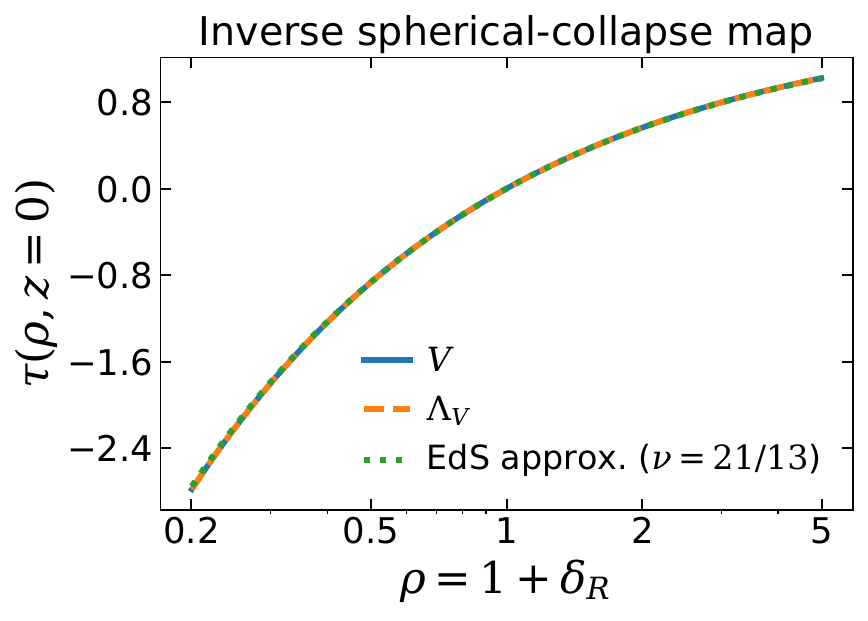}
\hfill
 \includegraphics[width=0.485\textwidth]{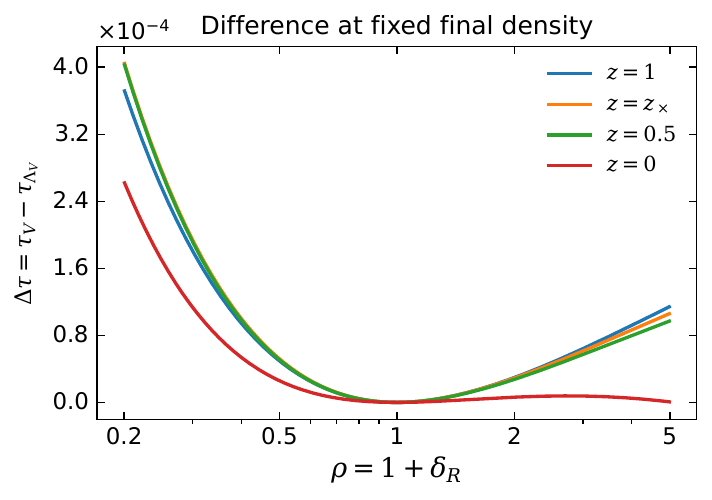}
 \caption{Inverse spherical-collapse maps, computed with standard
 gravity. \textbf{Left:} the linear density contrast $\tau(\rho)$ at
 $z=0$ for $V$ (blue solid), $\Lambda_V$ (orange dashed) and the
 Einstein--de Sitter approximation of Ref.~\cite{Bernardeau:1994} (green
 dotted). \textbf{Right:} the difference
 $\Delta\tau\equiv\tau_V-\tau_{\Lambda_V}$ at $z=1$ (blue), $z_\times$
 (orange), $z=0.5$ (green) and $z=0$ (red). We show the difference rather
 than the ratio because $\tau$ changes sign and vanishes at $\rho=1$,
 where $\Delta\tau=0$ by construction.}
 \label{fig:collapse-map-v9}
\end{figure*}

Spherical collapse is one of the main ingredients of the theoretical prediction for the PDF presented in Sec.~\ref{sec:ldt}, which we compare with the simulations in Sec.~\ref{sec:ldt-results}. It provides the map between the linear density contrast $\tau$ and the final density $\rho$. We therefore first check how much the F42 expansion history changes this map. Figure~\ref{fig:collapse-map-v9} shows the map and its response. The left panel shows the map at $z=0$ for $V$ and $\Lambda_V$, together with the EdS approximation of Ref.~\cite{Bernardeau:1994}. A negative linear density contrast produces an underdensity, a positive one an overdensity, and $\tau=0$ gives $\rho=1$. The maps are very similar. Over the displayed interval $0.2\leq\rho\leq5$, $V$ differs from the EdS approximation by less than $1\%$; this comparison includes the error of the EdS approximation and does not isolate the effect of changing cosmology. To resolve the small difference between $V$ and $\Lambda_V$, the right panel shows the difference $\tau_V-\tau_{\Lambda_V}$ at all four epochs, which avoids dividing by zero at the mean density. The magnitude of the fractional difference $\tau_V/\tau_{\Lambda_V}-1$ remains below $0.03\%$ for $0.05\leq\rho\leq20$ at all epochs, excluding $\rho=1$, where the ratio is undefined because both contrasts vanish. An independent numerical integration reproduces the inverse maps to an absolute accuracy of order $10^{-8}$, so this small difference is numerically resolved.

The sign of the difference must be read with care. On the overdense side, a positive difference means that $V$ needs a slightly larger linear density contrast to reach the same density. On the underdense side, it means that a slightly less negative linear density contrast is needed. The difference therefore does not correspond to a uniform enhancement of density or gravity. Comparing instead the final densities reached from the same linear density contrast gives larger differences near collapse,
where the density grows steeply with $\tau$. The two comparisons must therefore be made at matched variables; checks of the inverse and forward mappings are discussed in Appendix~\ref{app:new-collapse-check}.

These tests indicate that the PDF response is not primarily driven by the small change in the spherical-collapse mapping. As we show next, it follows mainly the change in the width of the distribution.

\subsection{Variance and PDF response}
\label{sec:Variance-PDF-response}

\begin{figure*}[t]
 \centering
 \includegraphics[width=0.485\textwidth]{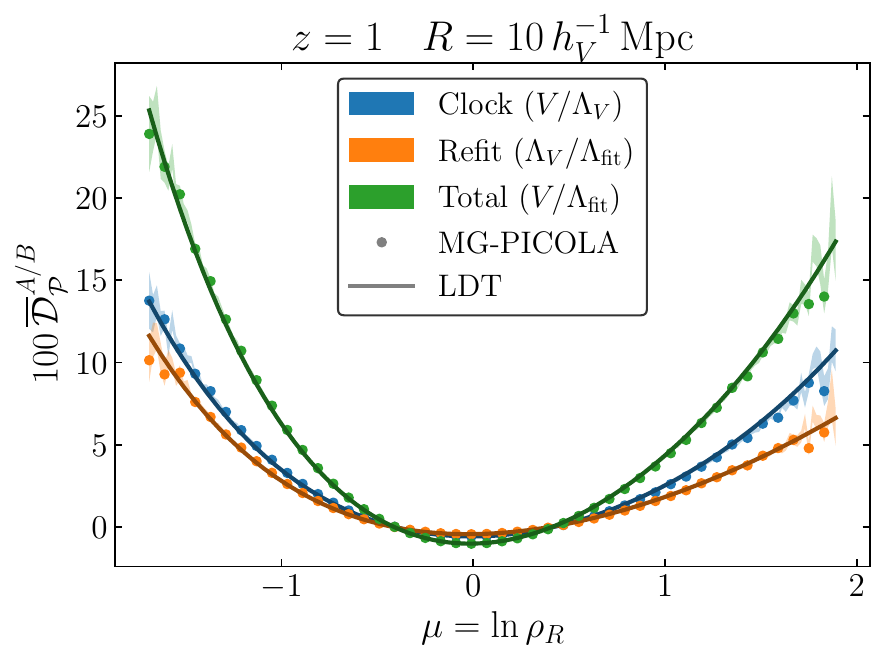}
\hfill
 \includegraphics[width=0.485\textwidth]{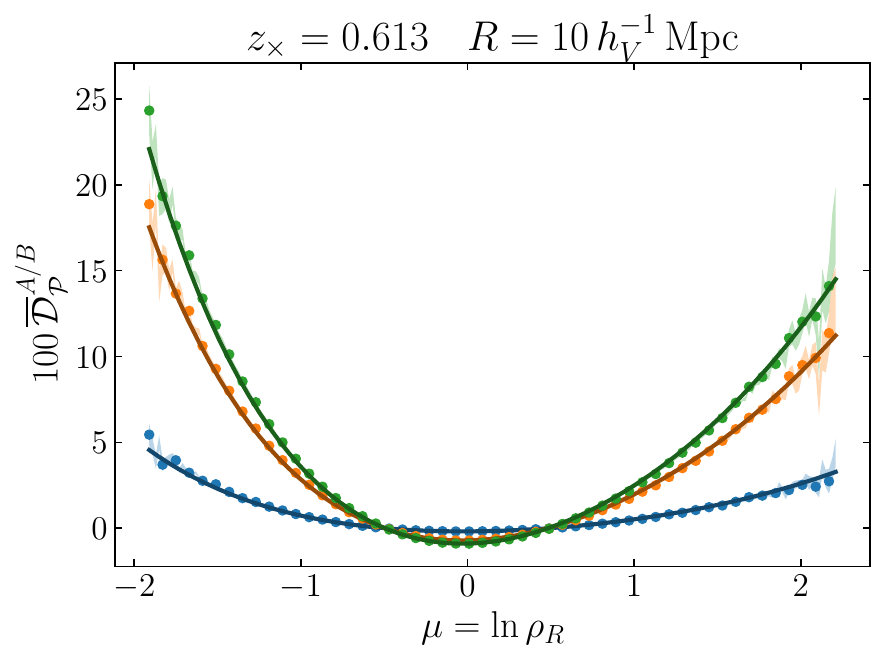}
\\[4pt]
 \includegraphics[width=0.485\textwidth]{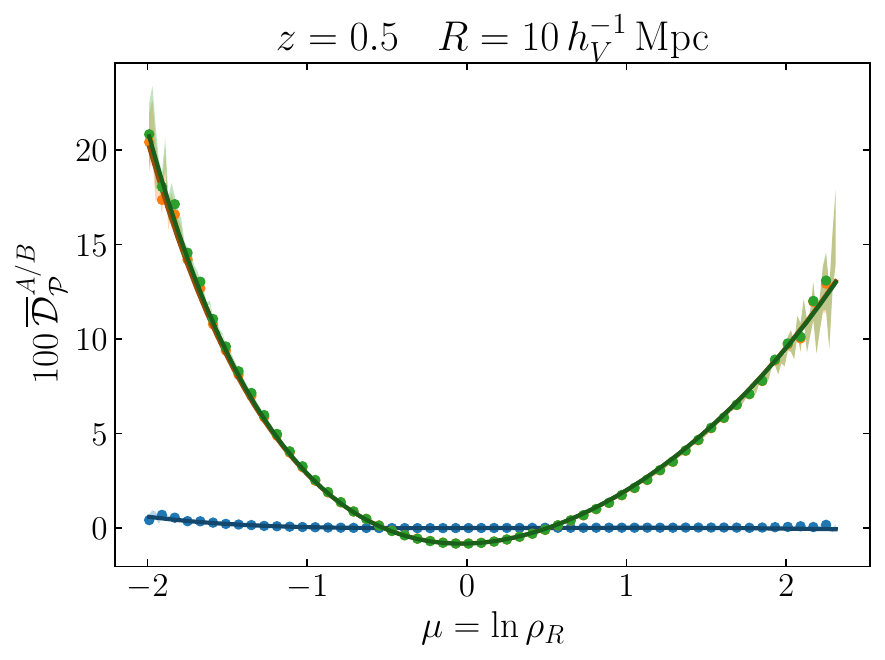}
\hfill
 \includegraphics[width=0.485\textwidth]{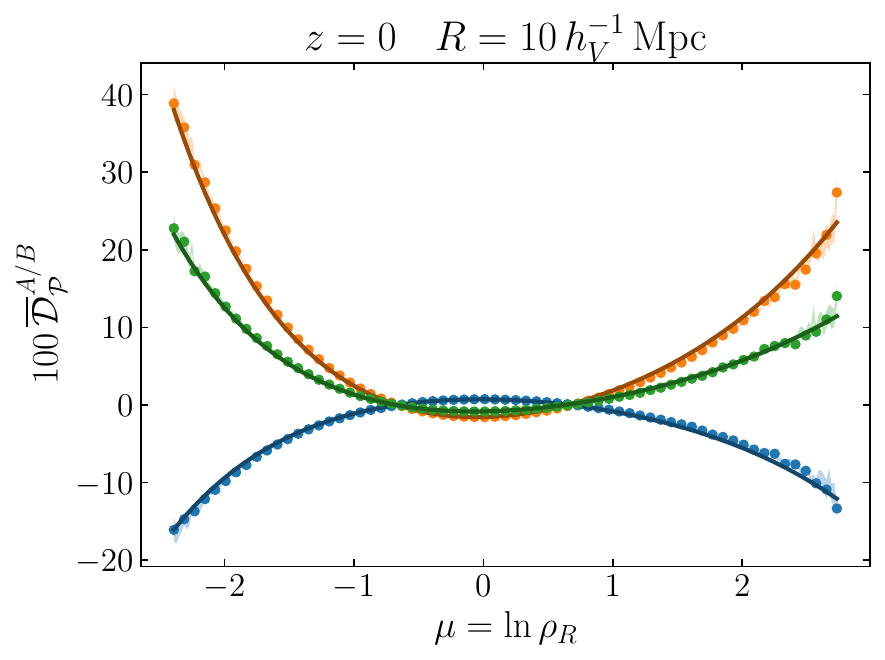}
\caption{Responses of the log-density PDF, $\Pc_\mu$, at $R=10\,h_V^{-1}\,{\rm Mpc}$ for the clock ($V$ relative to $\Lambda_V$, blue), refit ($\Lambda_V$ relative to $\Lambda_{\rm fit}$, orange) and total ($V$ relative to $\Lambda_{\rm fit}$, green) comparisons, at $z=1$, $z_\times=0.613$, $0.5$ and $0$. The plotted quantity is $100\,\overline{\mathcal{D}}_{\mathcal{P}}^{A/B}=100\,\langle\ln(\Pc_{\mu,A}/\Pc_{\mu,B})\rangle$, the mean over the 20 paired realizations of the log ratio of the PDFs of models $A$ and $B$. Points show the MG-PICOLA measurements, with shaded bands for 95\% paired-bootstrap intervals, and solid lines the simulation-calibrated LDT prediction.}
 \label{fig:triplet-pdf-v9e}
\end{figure*}

We next examine the one-point density PDF in spheres of comoving radius $R$. Figure~\ref{fig:triplet-pdf-v9e} shows the clock, refit and total log responses at $R=10\,h_V^{-1}\,\mathrm{Mpc}$, together with the LDT prediction of Sec.~\ref{sec:ldt}. The $\Lambda_{\rm fit}$ cells are evaluated at the same comoving radius in Mpc as the other suites (Table~\ref{tab:simulation}); the remaining radii appear in Appendix~\ref{app:pdf-radius-panels}. The plotted quantities are the seed-averaged log ratios $100\mDev_{\Pc}$, and the shaded bands show their 95\% paired-bootstrap interval over the 20 seeds. To avoid noisy ratios in the tails, the figures show only bins containing at least 100 cells in every suite and realization. The quantitative results, i.e. the largest PDF responses in Table~\ref{tab:response} and the rms norms below, use only the bins within the central $98\%$ of the
probability.

The clock response of $\Pc_\mu$, the PDF of the log-density $\mu=\ln\rho_R$, is driven mainly by a change in its width. We first compare the responses of the density variance and the log-density variance at fixed smoothing scale (third and fourth columns of Table~\ref{tab:response}). The density variance, $\Var(\rho_R)$, follows the power-spectrum response on the scales probed by the smoothing window. The response of the log-density variance, $\Var(\mu)$, is about $15\%$ smaller in magnitude at $z=1$ and $z_\times$, and about $25\%$ smaller today, at $R=10\,h_V^{-1}\,\mathrm{Mpc}$. The difference decreases at larger radii. Taking the logarithm compresses the high-density tail, reducing the weight of very dense cells. While $\Var(\mu)$ sets the width of $\Pc_\mu$, the shape of the PDF also depends on its skewness and higher moments~\cite{Uhlemann:2016}.

In the clock comparison (blue points in Fig.~\ref{fig:triplet-pdf-v9e}), the PDF records how much structure has grown. Since $V$ and $\Lambda_V$ have the same background matter density at all times, their expansion differs only through the effective dark energy of F42. Starting from the same early fluctuation amplitudes, $V$ first expands more slowly than $\Lambda_V$, since its phantom effective dark energy is less dense than a cosmological constant at early times. The smaller $H$ increases $\Omega_m(a)$ and changes the Hubble friction in the growth equation. As a result, matter in $V$ clusters more efficiently at early times. From  $z\simeq1.5$ onwards, the effective dark energy of $V$ becomes denser than a cosmological constant, with the largest excess at the phantom crossing, so $V$ expands faster and its growth slows down. At $z=1$ and at the crossing, $V$ still keeps part of its early advantage. Near $z=0.5$ the advantage has been erased, and the two distributions have almost the same width. Today it has turned into a deficit, so $V$ has fewer extreme cells than $\Lambda_V$ and the pattern is reversed. The PDF thus reflects the accumulated growth (Sec.~\ref{sec:background-results}).

The comparisons with $\Lambda_{\rm fit}$ include the shift in the cosmological parameters. In the refit comparison (orange points), $\Lambda_V$ and $\Lambda_{\rm fit}$ share the $\Lambda$CDM background prescription, $w=-1$, but their different fitted parameters give different $H(a)$ and linear power spectra. At the common comoving scales of our cells, the combined parameter shift gives $\Lambda_V$ more linear power (Sec.~\ref{sec:Power-response}), so structure is more developed and its distribution is broader at all epochs. The total comparison (green points), $V$ relative to $\Lambda_{\rm fit}$, combines this parameter effect with the expansion history, since its log ratio is the sum of the clock and refit ones [Eq.~\eqref{eq:clock-refit-log}]. At $z=1$, both effects widen the distribution of $V$, and the expansion history contributes most. By the crossing, as growth in $V$ slows down, the parameter shift has taken over, and near $z=0.5$, where the clock effect is very small, the total and refit curves nearly coincide. Today the slower late growth of $V$ reduces the spread of its densities, partly cancelling the effect of the parameter shift, so the distribution of $V$ remains wider than that of $\Lambda_{\rm fit}$, but by less than in the refit comparison.

To compare the shapes of the PDFs independently of their means and widths, we use the standardized log-density
$y_M=(\mu-\langle\mu\rangle_M)/\sigma_{\mu,M}$, where $\sigma_{\mu,M}$ is the standard deviation of $\mu$ in model $M$ (Appendix~\ref{app:standardized-pdf}, Fig.~\ref{fig:standardized-pdf}). This greatly reduces the clock response at $z=1$ and $z=0$, where the change in variance is largest. LDT reproduces this reduction and the main features of the remaining response. Its sign-changing, roughly antisymmetric pattern near the centre suggests a contribution from skewness. The direct measurements in Appendix~\ref{app:standardized-pdf}, however, show that skewness alone does not explain the differences between LDT and the simulations. These differences may also reflect higher moments, the approximations in LDT and COLA, and histogram reconstruction.

\subsection{Simulation-calibrated LDT prediction}
\label{sec:ldt-results}

The LDT prediction (solid lines in Fig.~\ref{fig:triplet-pdf-v9e}) is calibrated on the measured log-density variance of each model. Consequently, the difference in the widths of the two distributions is supplied by the simulations rather than independently predicted. However, the density dependence of the PDF response is not imposed by this calibration, but follows from the LDT  construction, including the linear power spectrum and spherical-collapse dynamics.

For the clock comparison, the predicted PDF ratio differs from the measured ratio by about $0.25\%$ or less in every retained central bin, at all four redshifts and smoothing radii. More precisely, the difference between their log ratios is smaller than $2.5\times10^{-3}$ in magnitude. Since the clock effect is itself small, it is useful to compare this error with the size of the effect.

To quantify how well LDT reproduces the clock response, we summarize the comparison with two numbers: the typical size of the measured response,
$A_{\rm rms}$, and the typical size of the LDT error, $E_{\rm rms}$, so that their ratio gives the error as a fraction of the response. Both are computed over the retained bins $i\in\mathcal S$ within the central $98\%$ of the probability distribution. We weight each bin by its probability so that bins with many cells count more than the sparse tails. The weights are
\begin{equation}
 w_i=\frac{\bar p_{V,i}+\bar p_{\Lambda_V,i}}
 {\sum_{j\in\mathcal S}(\bar p_{V,j}+\bar p_{\Lambda_V,j})},
 \label{eq:rms-weights}
\end{equation}
where $\bar p_{M,i}$ is the probability of bin $i$ in model $M$, averaged
over the 20 realizations. With
$\Dev_{\Pc}^{\rm LDT}(\mu_i)\equiv\ln[\Pc^{\rm LDT}_{\mu,V}(\mu_i)/\Pc^{\rm LDT}_{\mu,\Lambda_V}(\mu_i)]$
the predicted log ratio of the PDFs, the two numbers are the
probability-weighted root mean squares
\begin{align}
 A_{\rm rms}^2 &\equiv \sum_{i\in\mathcal S}w_i\,\big[\mDev_{\Pc}(\mu_i)\big]^2,\nonumber\\
 E_{\rm rms}^2 &\equiv \sum_{i\in\mathcal S}w_i\,\big[\Dev_{\Pc}^{\rm LDT}(\mu_i)-\mDev_{\Pc}(\mu_i)\big]^2,
 \label{eq:rms-definition}
\end{align}
computed separately at each redshift and radius.

For example, at $z=1$ and $R=10\,h_V^{-1}\,\mathrm{Mpc}$, the typical clock response is $A_{\rm rms}\simeq6.5\times10^{-3}$ (about $0.65\%$), while the typical error is $E_{\rm rms}\simeq2.2\times10^{-4}$ (about $0.022\%$). Thus the error is about $3.4\%$ of the response, not $3.4\%$ of the PDF ratio. Here and below, percentages obtained by multiplying small log ratios by 100 agree with fractional percentages to leading order.

At $R=10\,h_V^{-1}\,\mathrm{Mpc}$, the ratio $E_{\rm rms}/A_{\rm rms}$ ranges from $3.4\%$ to $7.2\%$ (Table~\ref{tab:response}). It generally decreases at larger radii, reaching $1\%$ at $z=1$ and $R=25\,h_V^{-1}\,\mathrm{Mpc}$. The absolute error is largest at $z=0$, when the density field is most nonlinear. Near $z=0.5$, the absolute error is smallest but the response is almost zero, giving the largest relative error at $R=10\,h_V^{-1}\,\mathrm{Mpc}$.
These error estimates compare the PDFs at the same log-density $\mu$ (before standardization). They do not quantify the accuracy of the much smaller response after standardization. We discuss that comparison, including a direct measurement of skewness, in Appendix~\ref{app:standardized-pdf}.

An independent numerical reconstruction using separately evaluated spherical-collapse derivatives, together with the archived linear-variance grid and measured log-density variance calibration, changes the clock log-PDF response at fixed $\mu$ by less than $4.4\times10^{-6}$ on the retained central bins for all four epochs and four smoothing radii (Appendix~\ref{app:validation}). The adopted density-rescaling operation is also reproduced by the public \texttt{pyLDT} normalization kernel. These are implementation-consistency checks, not validation of the full F42 cosmological pipeline or of COLA dynamics.

\section{Robustness and interpretation}
\label{sec:discussion}

The preceding results distinguish three questions: the response to changing the expansion history at fixed cosmological parameters, the impact of independently refitting those parameters, and the ability of LDT to reproduce the PDF response after its width has been calibrated. Here we summarize what these comparisons establish and the physical and
numerical assumptions underlying their interpretation.

\subsection{What the comparison does and does not establish}

The comparison of $V$ with $\Lambda_V$ isolates the effect of changing the expansion history at fixed parameters, initial phases and particle force. The response follows the accumulated growth: neither the phantom crossing nor the equal Hubble rates today fix the epoch at which it changes sign. 
The comparison of $V$ with $\Lambda_{\rm fit}$ combines this effect with the shift in the fitted parameters. The two contributions add exactly in the log responses [Eq.~\eqref{eq:clock-refit-log}], so this comparison is not a measurement of the clock effect alone. In the PDF, the clock dominates the total response at $z=1$, while the parameter refit dominates from the phantom crossing onwards.

Finally, the LDT prediction is calibrated on the measured log-density variance of each model, obtained from the same simulations used to measure the PDF. It therefore predicts how the PDF ratio varies with density for the measured widths; it does not independently predict the widths themselves. The same realizations supply the variances and test the PDF prediction. Within this scope, it reproduces the clock, refit and total responses, including the asymmetry between underdense and overdense cells, with small residuals that do not imply exact agreement in every bin.

\subsection{Limitations and scope}

Our interpretation is subject to the following limitations: 
\begin{enumerate} 
 \item Equation~\eqref{eq:geff-finite-w} is a finite-wavelength  expression for the source coefficient of the reduced single-dust  perturbation equation, not itself a strong-subhorizon expansion.  Together with the friction coefficient in  Eq.~\eqref{eq:friction-exact}, it approaches the standard-growth form at  leading order for $\mathcal K\gg1$. The simulations adopt the corresponding weak-field Newtonian particle-force prescription, which is  not an all-orders derivation of nonlinear VCDM dynamics. Radiation is  retained in the numerical backgrounds and initialization but omitted  from the analytic single-dust identities.

 \item The numerical checks cover the time stepping of the power  response for one seed at $z=0$ (Appendix~\ref{app:validation}) and a  spatial-resolution test in a smaller box  (Appendix~\ref{app:mesh-ref}), which preserves the qualitative PDF  response but does not establish quantitative convergence at the  precision of the main ensemble.

 \item The LDT prediction uses the measured $\Var(\ln\rho_R)$ in each cosmology. We use the same realizations to test the PDF shapes. A test on a separate set of realizations would be useful, but would still not constitute an independent prediction of the nonlinear variance.

 \item We adopt massless neutrinos in all cosmologies, $N_{\rm ncdm}=0$.  This is a controlled numerical choice, not a physically complete  neutrino treatment. Extending the calculation to massive neutrinos would  require consistently treating their effects on the background,
 scale-dependent growth, and the matter field sampled by the simulations.

 \item We study the matter-density PDF in  spheres in real space, without  galaxy or halo bias, redshift-space distortions, baryonic hydrodynamics,  survey geometry, or an observational covariance. We therefore make no  claim of survey detectability or of additional parameter-constraining  power.
 
\item Our simulations use the COLA approximation, which has not been validated against full $N$-body simulations for the one-point PDF in these cosmologies. A shared numerical method and resolution can promote partial error cancellation, but do not establish convergence of the responses. The smaller-box resolution pilot preserves their qualitative behaviour, but shows numerical sensitivities exceeding the production sampling precision in several diagnostics; it does not determine a correction to the production volume. The absolute PDF, calibrated variance and small standardized residual therefore remain subject to numerical systematics.
A matched full-$N$-body comparison is left for future work.
\end{enumerate}

\section{Conclusions}
\label{sec:conclusions}

We have quantified how the phantom-crossing F42 expansion history of VCDM affects the nonlinear matter distribution, using matched simulations that isolate the expansion history ($V$ versus $\Lambda_V$) and that include the shift between the two best fits ($V$ versus $\Lambda_{\rm fit}$). Our main results are:
\begin{enumerate}
 \item The expansion history leaves an accumulated, sign-changing imprint on structure growth. Starting from equal amplitudes at $z=200$, growth  in $V$ is enhanced at $z=1$, nearly equal to that in $\Lambda_V$ near  $z=0.5$, and suppressed today, so the sign change occurs well after the  phantom crossing. The matter power spectrum follows this sequence, from  about $+1.2\%$ at $z=1$ to $-1.5\%$ today.

 \item The density PDF shows the same reversal, mainly through the change  in its width: its response is U-shaped when $V$ has the broader  distribution and inverted today, and is larger in underdense than in  overdense cells. The spherical-collapse map, by contrast, changes by less than $0.03\%$ over the tested density range, excluding $\rho=1$, where the ratio is undefined.

 \item Between the two best fits, the differences in the fitted cosmological parameters dominate the  difference in the power spectrum at all epochs, with the expansion  history visible on top of it. In the PDF, the expansion history  dominates at $z=1$ and the cosmological parameter differences from the crossing onwards.

 \item Calibrated on the measured log-density variance of each model,
LDT broadly reproduces the PDF responses at fixed $\mu$ in all three comparisons.
For the controlled comparison at fixed $\mu$ on the retained central bins,
its probability-weighted residual is approximately $1$--$7\%$ of the
response amplitude. This accuracy statement does not apply to the much
smaller response after standardization in Appendix~\ref{app:standardized-pdf}:
LDT reproduces the main features, with differences that could arise from approximations in LDT and COLA or from histogram reconstruction.
\end{enumerate}

The PDF response mainly reflects how much structure has grown: removing the differences in the mean and width of the log-density PDF leaves a much smaller shape difference at the epochs with the largest variance responses. A change in width alone can give a U-shaped response, so that pattern does not by itself establish information beyond the variance. Within the Newtonian particle-force approximation, our results nevertheless show that a modified expansion history leaves a coherent, redshift-dependent imprint on the nonlinear matter distribution, even when the short-range force is unchanged.

\paragraph*{Outlook.}
These results motivate several next steps. First, we will validate our COLA results against full $N$-body simulations, both the responses between models and the absolute PDF and its variance at the smallest smoothing radius, where the approximate dynamics and the mesh resolution matter most. Second, larger simulation sets, including variations of the cosmological parameters and enough realizations to estimate covariances, will allow Fisher forecasts and simulation-based inference that quantify how much the PDF adds to the power spectrum in constraining phantom-crossing expansion histories. Finally, extending the analysis to tracers, with bias, redshift-space distortions and survey effects, and to massive neutrinos, will make it possible to confront these predictions with DESI data.

\begin{acknowledgments}
The present project originated from scientific discussions held during the Fourth Workshop on Current Challenges in Cosmology, hosted by Universidad del Valle in Cali, Colombia, in May 2026. A.D.F.~warmly thanks Alexander Gallego-Cadavid and J.~Bayron Orjuela-Quintana for their generous hospitality and for fostering the stimulating scientific environment in which this collaboration began. He also gratefully acknowledges the hospitality of the Department of Physics at Universidad del Valle. LC and CU were supported by the European Union (ERC StG, LSS BeyondAverage, 101075919). 

The authors acknowledge the use of Maple for computer algebra calculations, and of ChatGPT and Codex (OpenAI) for assistance with code development, numerical checks, drafting and proofreading. The authors have verified the scientific content and reviewed the text for accuracy. They take full responsibility for the results, conclusions and manuscript. None of these tools is an author.
\end{acknowledgments}

\appendix

\section{Best-fit parameters and likelihood contributions}
\label{app:likelihood-breakdown}

Table~\ref{tab:cosmology-full} lists the input parameters of the three
simulation suites at full precision. The values of $\sigma_8$ are not
input parameters but are derived from the linear spectra used to
initialise the simulations. They differ between $V$ and $\Lambda_V$,
despite the shared parameters, because the two models grow structure
differently from the same initial amplitude.

\begin{table}[h]
 \caption{Full-precision input parameters of the simulation suites. $H_0$
 is in units of $\mathrm{km\,s^{-1}\,Mpc^{-1}}$.}
 \label{tab:cosmology-full}
 \centering
 \begin{ruledtabular}
 \begin{tabular}{lccc}
 Parameter & $V$ & $\Lambda_V$ & $\Lambda_{\rm fit}$ \\
 \hline
 $H_0$             & 66.5676744   & 66.5676744 & 68.6730943 \\
 $\Omega_m$        & 0.32015328   & 0.32015328 & 0.29871835 \\
 $\Omega_b$        & 0.05068268   & 0.05068268 & 0.04773159 \\
 $10^9 A_s$        & 2.091066     & 2.091066   & 2.103402   \\
 $n_s$             & 0.968005     & 0.968005   & 0.9704057  \\
 $\tau_{\rm reio}$ & 0.0528345    & 0.0528345  & 0.0563673  \\
 $w_0$             & $-0.7872993$ & $-1.0$     & $-1.0$     \\
 \hline
 $\sigma_8$        & 0.81066198   & 0.81668496 & 0.82077884 \\
 \end{tabular}
 \end{ruledtabular}
\end{table}

Table~\ref{tab:likelihood-breakdown} lists the final unscaled
contributions at the separately optimized F42 and $\Lambda$CDM
likelihood minima. The quoted digits identify the numerical evaluations
and do not represent uncertainties on the minima.

\begin{table}[ht]
 \caption{Final unscaled likelihood contributions. The last column is F42
 minus $\Lambda_{\rm fit}$. Contributions are evaluated at the separately
 optimized parameter vectors; this is an objective breakdown, not a
 conditional or leave-one-dataset-out fit.}
 \label{tab:likelihood-breakdown}
 \centering
 \begin{ruledtabular}
 \begin{tabular}{lrrr}
 Block & F42 & $\Lambda_{\rm fit}$ & Difference \\
 \hline
 High-$\ell$ TTTEEE & 2344.6120 & 2346.8501 & $-2.2381$ \\
 Low-$\ell$ EE & 395.7588 & 396.2735 & $-0.5147$ \\
 Low-$\ell$ TT & 22.9660 & 22.5754 & $+0.3906$ \\
 CMB lensing & 9.0698 & 8.8542 & $+0.2156$ \\
 DESI DR2 BAO & 7.7764 & 10.9233 & $-3.1469$ \\
 \hline
 Total & 2780.1830 & 2785.4765 & $-5.2935$ \\
 \end{tabular}
 \end{ruledtabular}
\end{table}

\section{Spherical-collapse maps: initial conditions and numerical checks}
\label{app:new-collapse-check}

This appendix specifies the initial conditions for the spherical-collapse
integration and documents numerical checks of the resulting inverse maps
$\tau_M(\rho,z)$.

\paragraph{Initial conditions.}
We integrate Eq.~\eqref{eq:collapse} from $a_i=10^{-4}$, using an initial
linear contrast $\epsilon$ at $a_i$ and the initial conditions
\begin{align}
 \delta_i&=\epsilon+\frac{17}{21}\epsilon^2,\qquad
 \delta_i'=f_i\left(\epsilon+\frac{34}{21}\epsilon^2\right),
 \label{eq:collapse-ic}\\
 f_i&=\frac{(3/2)\,y_i}{1+(3/2)\,y_i},\qquad
 y_i=\frac{\Omega_{m,0}\,a_i}{\Omega_{r,0}}.
 \label{eq:meszaros-slope}
\end{align}
Here $f_i$ is the growth rate of the M\'esz\'aros growing mode in the
presence of radiation; for our parameters $f_i\simeq0.337$, well below
its EdS value of one. The second-order coefficients take their EdS values
and serve to suppress transients; since they are not exact during
radiation domination, we test the sensitivity of the maps to $a_i$ below.
The main maps are computed with a ninth-order Runge--Kutta (Vern9)
integrator.

\paragraph{Independent integration.}
As an independent check, we solve the collapse equation in terms of the
log-density $\eta=\ln(1+\delta)$ and $v=\eta'$,
\begin{equation}
 \eta'=v,\qquad
 v'=-\left(2+\frac{H'}{H}\right)v+\frac{v^2}{3}
    +\frac32\Omega_m\left(\e^{\eta}-1\right).
 \label{eq:logrho-integration}
\end{equation}
At fixed output redshift, let $J=\partial\eta/\partial\tau$ and $U=J'$,
where $\tau$ is the linear density contrast that labels the solution.
They obey
\begin{equation}
 J'=U,\qquad
 U'=\left[-\left(2+\frac{H'}{H}\right)+\frac23v\right]U
    +\frac32\Omega_m\e^{\eta}J.
 \label{eq:collapse-tangent}
\end{equation}
For each target density, we solve $\eta(a_f;\tau)=\ln\rho$ by shooting,
starting from the Bernardeau approximation rather than from the main
maps, and using an eighth-order Runge--Kutta (DOP853) integrator with the
initial conditions of Eq.~\eqref{eq:collapse-ic}. This recovers all
entries of the maps (701 densities, four epochs and two models), with a
maximum residual in the final log-density of $6.2\times10^{-11}$. The
maximum difference from the main maps is $8.1\times10^{-9}$, which
includes differences in the interpolation of the background and in the
shooting. Tightening the integration tolerance changes the results by
less than $1.2\times10^{-10}$ at the tested densities.

\paragraph{Second derivative and rate curvature.}
The prefactor of Eq.~\eqref{eq:raw-pdf} requires second derivatives of
the rate function. We evaluate them in $\mu=\ln\rho$, using
\begin{equation}
 \Psi''(\rho)+\frac{\Psi'(\rho)}{\rho}
 =\rho^{-2}\frac{\dd^2\Psi}{\dd\mu^2}.
 \label{eq:curvature}
\end{equation}
To obtain the collapse contribution to this curvature without taking
second differences of an interpolated map, we also integrate
$Q=\partial^2\eta/\partial\tau^2$,
\begin{align}
 Q''&=\left[-\left(2+\frac{H'}{H}\right)+\frac23v\right]Q'\nonumber\\
&\quad+\frac23(J')^2+\frac32\Omega_m\e^{\eta}\left(Q+J^2\right),
 \nonumber\\
 \frac{\dd\tau}{\dd\mu}&=\frac1J,\qquad
 \frac{\dd^2\tau}{\dd\mu^2}=-\frac{Q}{J^3}.
 \label{eq:second-tangent}
\end{align}
This checks the collapse contribution to the rate curvature; the
linear-variance derivatives and the nonlinear width enter the PDF
separately.

\paragraph{Sensitivity to the initial time.}
Changing $a_i$ to $5\times10^{-5}$ or $2\times10^{-4}$, at $z=0$ and $1$
and at densities $\rho=0.05$, 0.1, 0.2, 0.5, 1, 2, 5, 10 and 20, changes
the difference $\tau_V-\tau_{\Lambda_V}$ by at most $6.3\times10^{-7}$ for
$0.2\leq\rho\leq5$, and by up to $5.5\times10^{-6}$ in the most
underdense case, $\rho=0.05$. This tests the initial conditions,
independently of the integration accuracy. The same calculation gives
$D_V(0)/D_{\Lambda_V}(0)=0.993$, with the ratio crossing unity near
$z\simeq0.5$ when both growth factors are normalized to agree at
$z=200$.

\paragraph{Collapse thresholds.}
As an additional diagnostic, we compute the linear threshold for formal
collapse, $\delta_c$, by integrating $y=(1+\delta)^{-1/3}$ down to
$y_{\min}=10^{-5}$ with the initial conditions of
Eq.~\eqref{eq:collapse-ic}:
\begin{center}
\begin{tabular}{lrr}
\toprule
Epoch & $\delta_{c,V}$ & $\delta_{c,\Lambda_V}$\\
\midrule
$z=1$ & 1.68427 & 1.68386\\
$z=z_\times$ & 1.68268 & 1.68240\\
$z=0.5$ & 1.68191 & 1.68172\\
$z=0$ & 1.67543 & 1.67578\\
\bottomrule
\end{tabular}
\end{center}
For comparison, the same calculation in EdS gives $1.6864702$. Changing
$y_{\min}$ from $10^{-4}$ to $10^{-5}$ shifts the thresholds by about
$1.1\times10^{-6}$, while changing $a_i$ shifts them by up to
$4.5\times10^{-5}$, so the last digits reflect our implementation rather
than a general precision. These thresholds are not used in the PDF
calculation, which relies on the full maps $\tau(\rho,z)$.

\section{Density-field estimators and numerical validation}
\label{app:validation}

This appendix specifies the reconstruction of the smoothed matter-density
field and the statistical estimators used in Sec.~\ref{sec:estimators},
followed by numerical consistency and stability checks. These tests
validate the implementation; they do not provide an independent
high-resolution nonlinear $N$-body benchmark.

\subsection{Eulerian cells, CIC filtering, and PDF estimators}
\label{app:estimator-check}

\paragraph{Eulerian density and smoothing.}
For a uniformly sampled Eulerian cell of comoving radius $R$, the
relative matter density at redshift $z$ is
\begin{align}
 \rho_R(\boldsymbol x,z)
 &=\frac{1}{\bar\rho_m(z)V_R}
 \int_{|\boldsymbol y-\boldsymbol x|<R}
 \rho_m(\boldsymbol y,z)\,\dd^3y,\nonumber\\
 V_R&=\frac{4\pi R^3}{3},
 \label{eq:continuous-cell}
\end{align}
where the density and volume are consistently expressed in comoving
units. The cell centers sample volume rather than particle or halo
locations. Equal-mass particles are deposited on a periodic
$N_g^3=1024^3$ mesh by cloud-in-cell (CIC) assignment, which is distinct
from counts-in-cells statistics. Writing $\Delta x=L/N_g$, the
Fourier-space amplitude window of the CIC assignment is
\begin{equation}
 A_{\rm CIC}(\boldsymbol k)
 =\prod_{j=1}^{3}\operatorname{sinc}^2(k_j\Delta x/2),
 \qquad \operatorname{sinc}u=\frac{\sin u}{u}.
 \label{eq:cic-window}
\end{equation}
The smoothed contrast is obtained using the spherical top-hat window
$W(kR)$ of Eq.~\eqref{eq:window}. The raw and CIC-deconvolved fields are
\begin{equation}
 \widetilde\delta_R^{(d)}(\boldsymbol k)
 =\widetilde\delta_g(\boldsymbol k)
  \frac{W(kR)}{A_{\rm CIC}(\boldsymbol k)^d},
 \qquad d=0,1,
 \label{eq:mesh-filter}
\end{equation}
for nonzero Fourier modes; the zero mode is set to zero. An inverse
Fourier transform gives $\rho_R^{(d)}=1+\delta_R^{(d)}$ at each grid
site. The $d=1$ field removes the CIC assignment factor for represented
modes, but does not restore missing modes, remove all aliases, or
establish force-resolution convergence. In particular, deconvolution
need not guarantee positive density values on a finite mesh.

\paragraph{Log-density PDF.}
For positive cells, define $\mu_R=\ln\rho_R$. If $n_i$ grid sites lie in
the log-density bin $[\mu_i,\mu_{i+1})$ of width $\Delta\mu$, the
probability and PDF estimators are
\begin{equation}
 \widehat p_i=\frac{n_i}{N_g^3},\qquad
 \widehat\Pc_{\mu,i}=\frac{\widehat p_i}{\Delta\mu},\qquad
 \widehat\Pc_{\rho}(\rho_i)=\frac{\widehat\Pc_{\mu,i}}{\rho_i},
 \label{eq:histogram-estimator}
\end{equation}
where $\rho_i=\exp[(\mu_i+\mu_{i+1})/2]$. The last identity is a
bin-center Jacobian conversion; a precise finite-bin prediction should
compare integrated probabilities. Underflow, overflow and nonpositive
cells are counted separately, and excluding the displayed tails does not
renormalize the histograms: $N_g^3$ remains the denominator. Log-density
moments use positive cells, so their interpretation requires checking
the nonpositive-cell fraction.

\paragraph{Variances and shot noise.}
The analysis stores Fourier-space and real-space density variances, as
well as $\Var(\ln\rho_R)$. For the discrete Poisson-noise prescription
used by the estimator, the filtered contribution is
\begin{equation}
 S_{R,d}^{\rm Pois}=\frac{1}{N_p}
 \sum_{\boldsymbol k\ne\boldsymbol0}^{\rm mesh}
 W^2(kR)A_{\rm CIC}^{2(1-d)}(\boldsymbol k),
 \label{eq:shot-discrete}
\end{equation}
where $N_p$ is the number of particles and the mesh-mode sum includes
both signs of each Fourier mode. For the counts-in-cells estimators, only
the real-space density variance receives this subtraction; the
power-spectrum estimator has a separate noise treatment. The subtraction
is an estimator convention rather than a model-independent removal of
discreteness effects from evolved, initially lattice-distributed
particles, so comparisons of raw and deconvolved fields remain useful.

\paragraph{Data-vector selection and uncertainty.}
The power block contains 57 $k$ bins at each of four epochs, giving 228
coordinates. The variance block has four smoothing radii, three variance
definitions and four epochs, giving 48 coordinates. For the PDF, 300
candidate log-density bins spanning $-3\le\mu\le3$ are examined at each
radius and epoch. The central-$98\%$ mask and the requirement of at least
100 cells per bin in each model and realization leave 1297 PDF
coordinates, for a total of 1573. The bootstrap resamples realization
pairs (or seed triplets), not cells. When propagating uncertainty in the calibrated LDT prediction, the measured log-density variances are re-evaluated and the widths recalibrated
within each paired draw. This is distinct from the measurement-only bands
in Figs.~\ref{fig:triplet-pdf-v9e} and \ref{fig:standardized-pdf}, where the
theoretical curves are held fixed.

\paragraph{Lognormal illustration of the width effect}

To illustrate how much of the PDF response is explained by the change in
width alone, we consider the unit-mean lognormal
model~\cite{Coles:1991if,Kayo:2001gu}, in which $\mu$ is Gaussian with
variance $u_M$ and mean $-u_M/2$, so that $\langle\rho\rangle=1$:
\begin{equation}
 \Pc^{\rm LN}_{\mu,M}(\mu)=\frac{1}{\sqrt{2\pi u_M}}
 \exp\!\left[-\frac{(\mu+u_M/2)^2}{2u_M}\right].
\end{equation}
For two models $A$ and $B$, the log ratio of their PDFs is a parabola in
$\mu$,
\begin{align}
 \ln\frac{\Pc^{\rm LN}_{\mu,A}(\mu)}{\Pc^{\rm LN}_{\mu,B}(\mu)}
 &= \frac{1}{2}\mu^2\left(\frac{u_A-u_B}{u_Au_B}\right)
 + \frac{1}{2}\ln\!\left(\frac{u_B}{u_A}\right)\nonumber\\
 &\quad - \frac{1}{8}(u_A-u_B),
 \label{eq:width-toy}
\end{align}
which opens upwards when $A$ has the broader distribution and downwards
when it has the narrower one. Since this model contains no information
beyond the variance, comparing it with the measured response shows what
the width alone explains. With the measured log-density variances of $V$
and $\Lambda_V$ at $z=1$ and $R=10\,h_V^{-1}\,{\rm Mpc}$, for example,
Eq.~\eqref{eq:width-toy} gives
$100\ln(\Pc^{\rm LN}_{\mu,V}/\Pc^{\rm LN}_{\mu,\Lambda_V})\approx-0.55$ at
$\mu=0$, in agreement with the measured value. The lognormal model cannot,
however, reproduce the asymmetry between underdense and overdense cells,
which reflects the non-Gaussian shape of $\Pc_\mu$ and is captured by the
LDT prediction.

\subsection{Independent consistency checks}

\paragraph{Background and linear growth.}
All linear inputs share $H_0$ to numerical precision in the comparison
of $V$ with $\Lambda_V$, and adopt massless neutrinos. The
radiation-aware background agrees with an independent numerical
evaluation to fractional precision $10^{-12}$, and independent
integrations of the linear growth agree to $2\times10^{-11}$. The growth
response changes sign near $z=0.495$, distinct from the normalization
imposed at $z=200$. Over the primary power-spectrum range, the growth
inferred from the \class{} spectra is scale independent to
$2\times10^{-5}$ for $z\le1$ and $6\times10^{-5}$ at $z=19$, supporting
the nearly scale-independent growth comparison entering
Eq.~\eqref{eq:rate}.

\paragraph{Linear variances and $\sigma_8$.}
The linear cell-variance integrals are reproduced by independent
Gauss--Legendre quadrature to $5\times10^{-10}$. The ratio of the input
values of $\sigma_8$ for $V$ and $\Lambda_V$ (Table~\ref{tab:cosmology-full})
agrees with $D_V(0)/D_{\Lambda_V}(0)$, as expected when the
same-parameter spectra differ mainly through scale-independent growth.

\paragraph{Estimator implementation.}
An independent NumPy implementation was tested against the production C
estimator on a controlled synthetic snapshot and a $16^3$ grid, with raw
and deconvolved filters. Density and log-density moments and the discrete
shot-noise sums agree to relative precision $2\times10^{-10}$.
Additional tests reproduce every histogram count, and check the
conversion between bin probabilities and PDF values, underflow and
overflow, periodic boundaries, Parseval consistency and unit mean. These
tests check the estimator implementation on controlled inputs; they do
not rerun the full cosmological ensemble.

\paragraph{Statistical analysis.}
An independent reconstruction of the 1573-dimensional mean log-response
vector and its covariance from the stored per-seed measurements
reproduces the reported results to floating-point precision. With only 20
matched realizations, the sample covariance has rank at most 19, and the
reported intervals are pointwise rather than simultaneous confidence
regions.

\paragraph{LDT prediction.}
The width calibration reproduces the target log-density variances
within $5\times10^{-12}$ on its archived quadrature. Independent continuous
quadrature at the same fixed widths gives a relative variance drift below
$9\times10^{-8}$ for Fig.~\ref{fig:standardized-pdf}, without a new fit. Recalculation with independently evaluated
collapse-map derivatives of Appendix~\ref{app:new-collapse-check}, using
the archived linear-variance grid, changes the rescaled PDF log
response by at most $4.33\times10^{-6}$. The rate function remains convex
in $\mu=\ln\rho$ over the tested support, with minimum curvature about
$0.90$. The raw integration support is $0.05\leq\rho\leq20$ and is
rescaled consistently with Eq.~\eqref{eq:tilted-pdf}, without cropping
back to the original interval. Extending the raw support to
$0.02\leq\rho\leq100$ provides an additional convergence check; the last
tested enlargement changes the response by at most $2.2\times10^{-8}$
in log units. The density-rescaling formula is also reproduced by the
public \texttt{pyLDT} normalization kernel; this checks that operation,
not its full cosmological pipeline.\footnote{An exponential mean-density
tilt, $\Pc\propto\e^{-c\rho}\widetilde\Pc$, was retained only as an
alternative numerical check. Unlike the adopted log-density translation,
it changes the centered shape of the log-density distribution.}

\subsection{Analysis controls}

No scale or realization was selected according to the signal amplitude
or the LDT agreement. The following controls were fixed before inspecting
the main ensemble:
\begin{itemize}
 \item Changing the primary power interval to $[0.02,0.15]$,
 $[0.03,0.15]$ or $[0.03,0.2]$ in $h_V\,\mathrm{Mpc}^{-1}$ changes the
 averaged log response by at most $0.034\%$.
 \item Removing any one matched realization pair changes an individual
 coordinate by at most $0.014\%$.
 \item Removing one smoothing radius from the average changes it by at
 most $0.044\%$.
 \item The maximum difference between the raw and deconvolved fields is
 $0.16\%$, reported as an analysis-choice effect rather than included in
 the sampling error.
 \item The bootstrap standard errors agree with the analytical paired
 standard errors at the expected finite-resample accuracy.
\end{itemize}

\subsection{Timestep check}

\begin{figure}[t]
 \centering
 \includegraphics[width=\columnwidth]{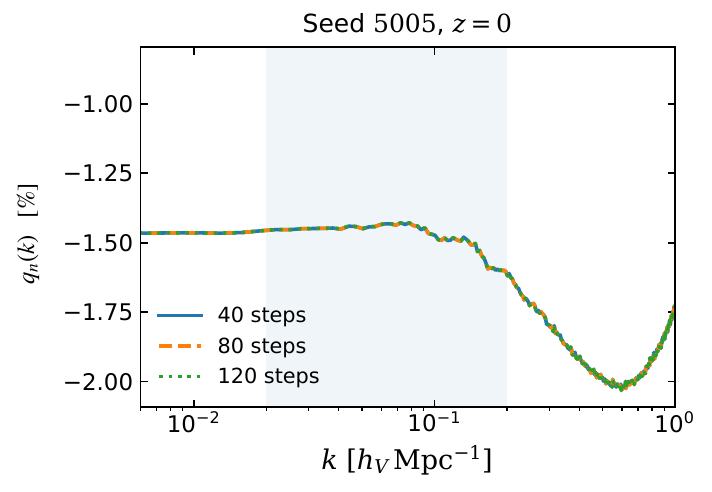}\\[2pt]
 \includegraphics[width=\columnwidth]{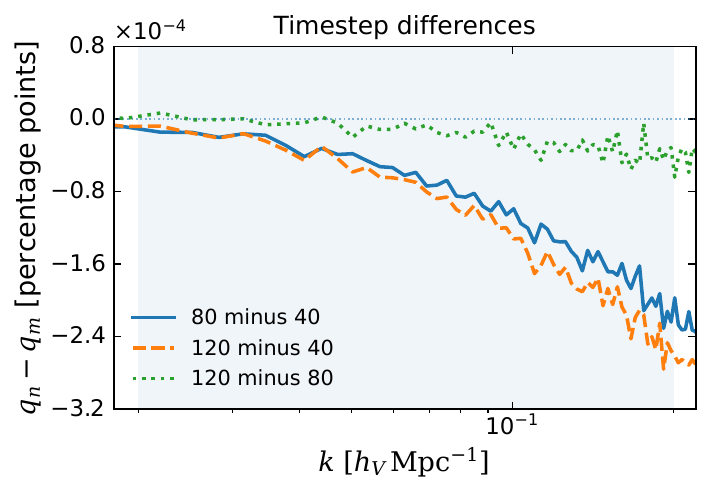}
 \caption{Timestep check for seed 5005 at $z=0$. Top: $q_n(k)$ for
 $n=40$, 80 and 120 timesteps, with the primary interval shaded. Bottom:
 the differences $q_{80}-q_{40}$, $q_{120}-q_{40}$ and $q_{120}-q_{80}$,
 in percentage points.}
 \label{fig:cola-convergence-v7}
\end{figure}

A dedicated six-run test uses the $V$ and $\Lambda_V$ parameters and 
input spectra, seed 5005 and $z=0$, with 40, 80 and 120 COLA timesteps.
These runs supplement, rather than replace, the main 20-pair ensemble.
We define the percentage power response at timestep count $n$ as
\begin{equation}
 q_n(k)=100\left[\frac{P_V^{(n)}(k)}{P_{\Lambda_V}^{(n)}(k)}-1\right],
 \label{eq:timestep-response}
\end{equation}
which differs from the ensemble log-response estimator of
Eq.~\eqref{eq:paired-deviation}. Over the 57 bins with
$0.02\le k/(h_V\,\mathrm{Mpc}^{-1})\le0.2$, the six spectra give
\begin{align}
 \max|q_{80}-q_{40}|&=2.31\times10^{-4},\nonumber\\
 \max|q_{120}-q_{40}|&=2.76\times10^{-4},\nonumber\\
 \max|q_{120}-q_{80}|&=5.55\times10^{-5},
 \label{eq:current-convergence-bounds}
\end{align}
in percentage points (Fig.~\ref{fig:cola-convergence-v7}).

The validation runs use a single interval from $z=19$ to $z=0$ with 40,
80 or 120 uniform steps in the scale factor, while the main runs use four
intervals ($19+5+2+14$ steps) to save the four epochs. The seed-5005
response from the main schedule differs from the single-interval 40-step
response by at most $1.97\times10^{-5}$ percentage points over the same
$k$ range, well below the percent-level clock signal. These checks assess
the time stepping of the relative power response for one seed at $z=0$;
they do not establish the convergence of the individual power spectra,
the PDF or other epochs. The spatial resolution is tested separately in
Appendix~\ref{app:mesh-ref}.

\subsection{Spatial-resolution test}
\label{app:mesh-ref}

As an additional spatial-resolution check, we performed 12 MG-PICOLA
simulations in a smaller box of side $1000\,h_V^{-1}\,\mathrm{Mpc}$,
varying the particle number and force-mesh resolution between $512^3$ and $1024^3$. The coarsest configuration has the cell size of the main simulations, about $2\,h_V^{-1}\,\mathrm{Mpc}$, and the finest reaches about $1\,h_V^{-1}\,\mathrm{Mpc}$, one tenth of our smallest smoothing radius. Despite some quantitative sensitivity to resolution, the clock PDF response keeps its qualitative behaviour in all configurations, including the U-shape at early times and its reversal at $z=0$. The limited number of realizations and the smaller box prevent this test from establishing quantitative convergence at the precision of the main ensemble.

\section{PDF responses at the other smoothing radii}
\label{app:pdf-radius-panels}

\begin{figure*}[t]
 \centering
 \includegraphics[width=0.24\textwidth]{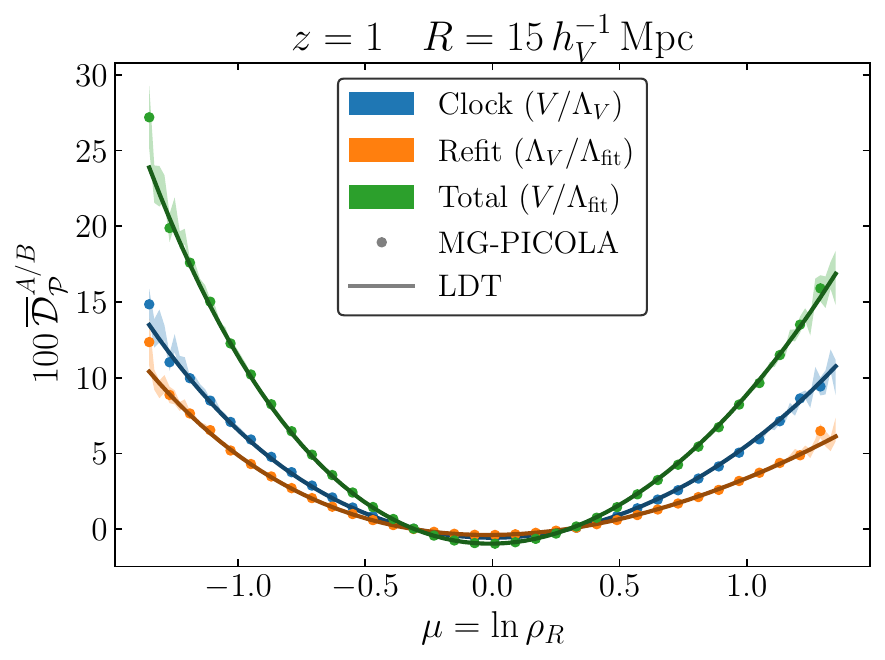}\hfill
 \includegraphics[width=0.24\textwidth]{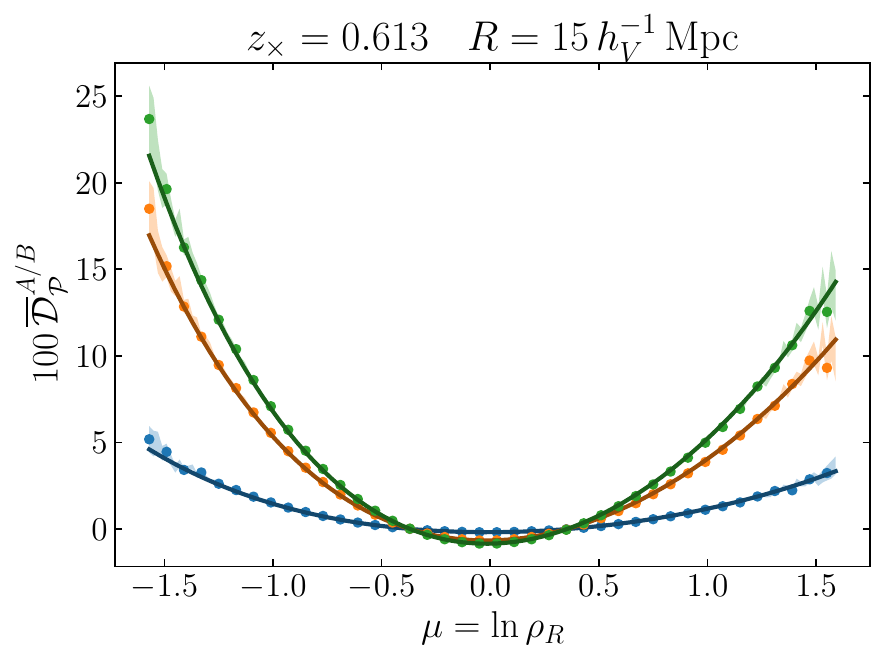}\hfill
 \includegraphics[width=0.24\textwidth]{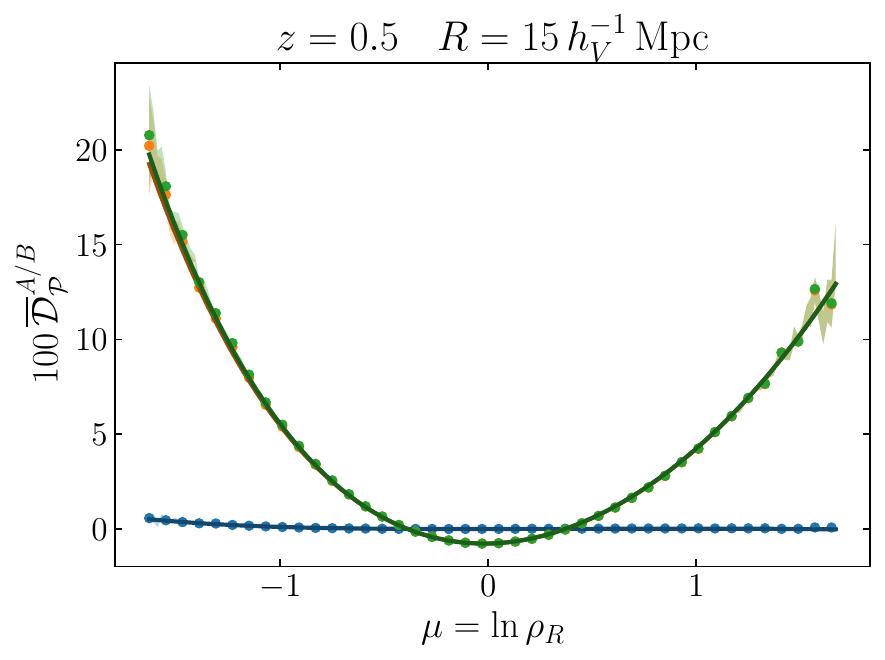}\hfill
 \includegraphics[width=0.24\textwidth]{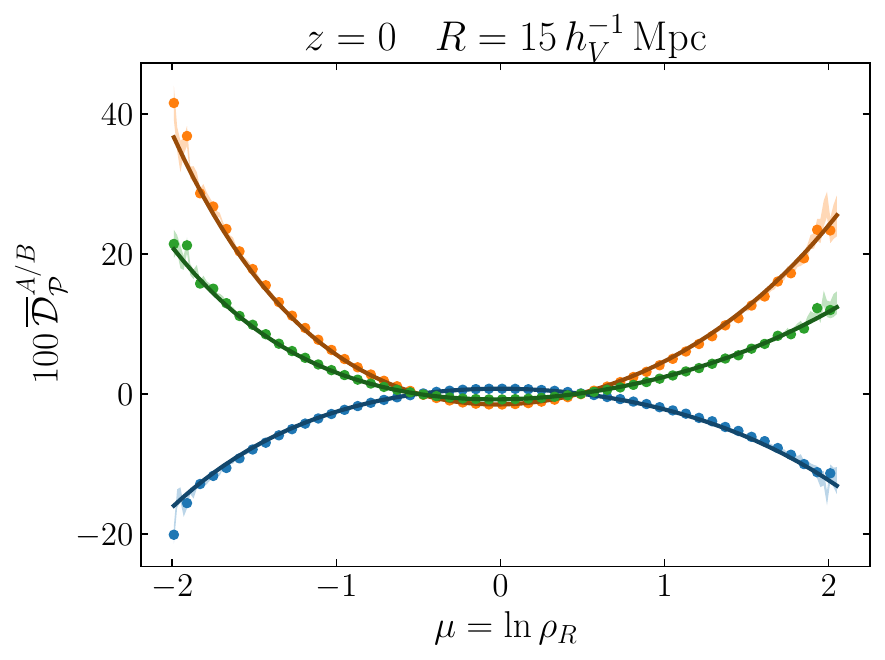}
 \\[6pt]
 \includegraphics[width=0.24\textwidth]{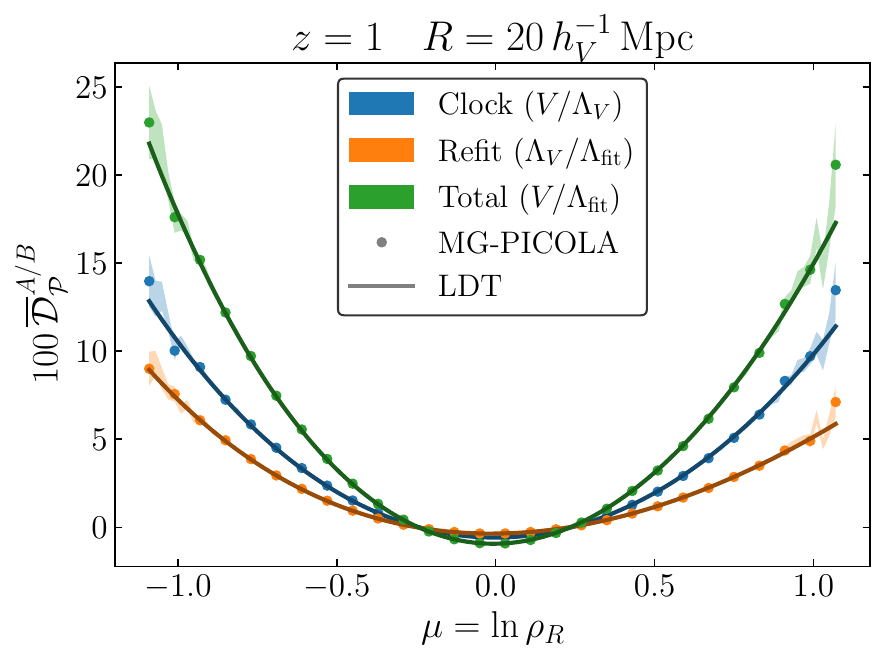}\hfill
 \includegraphics[width=0.24\textwidth]{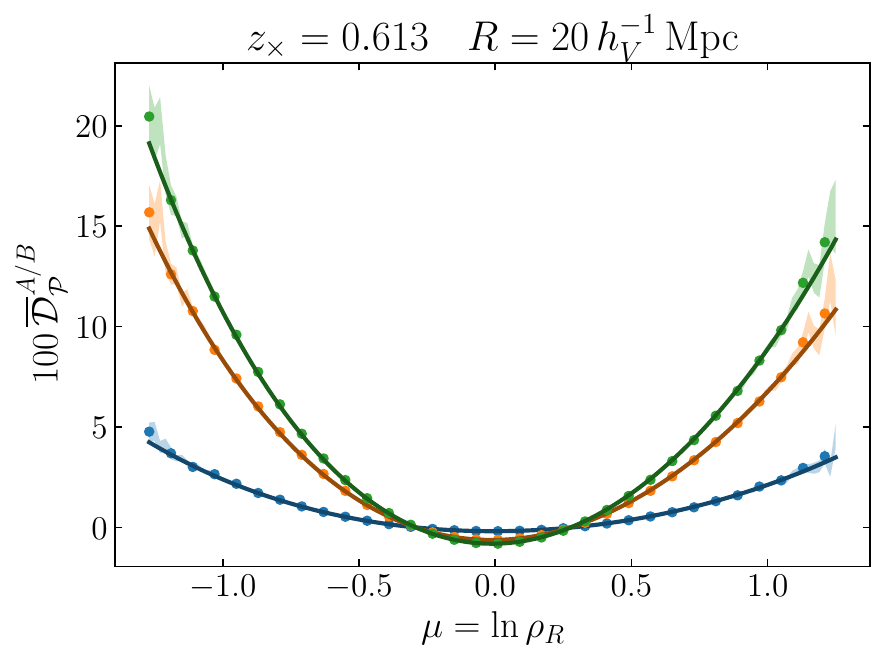}\hfill
 \includegraphics[width=0.24\textwidth]{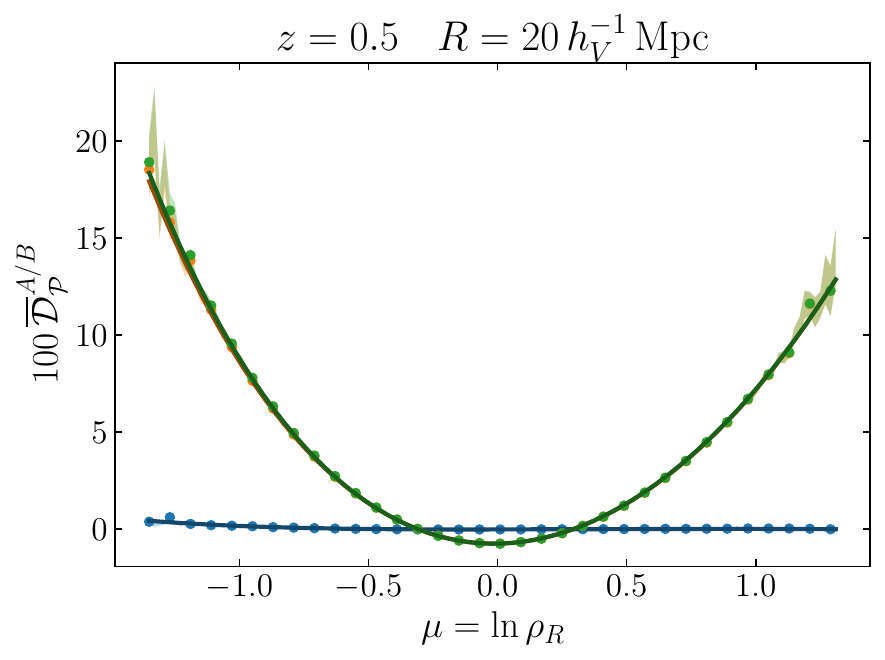}\hfill
 \includegraphics[width=0.24\textwidth]{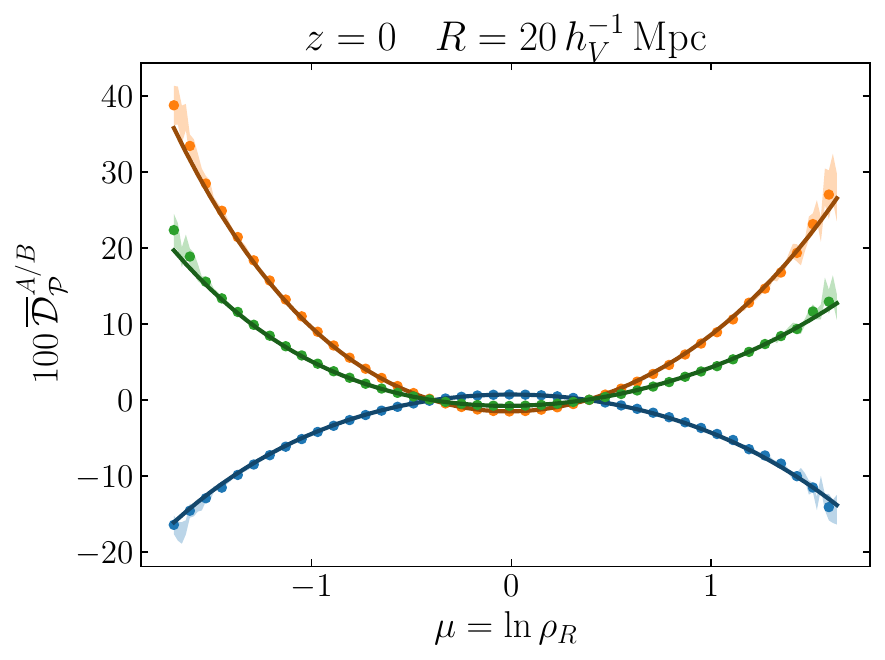}
 \\[6pt]
 \includegraphics[width=0.24\textwidth]{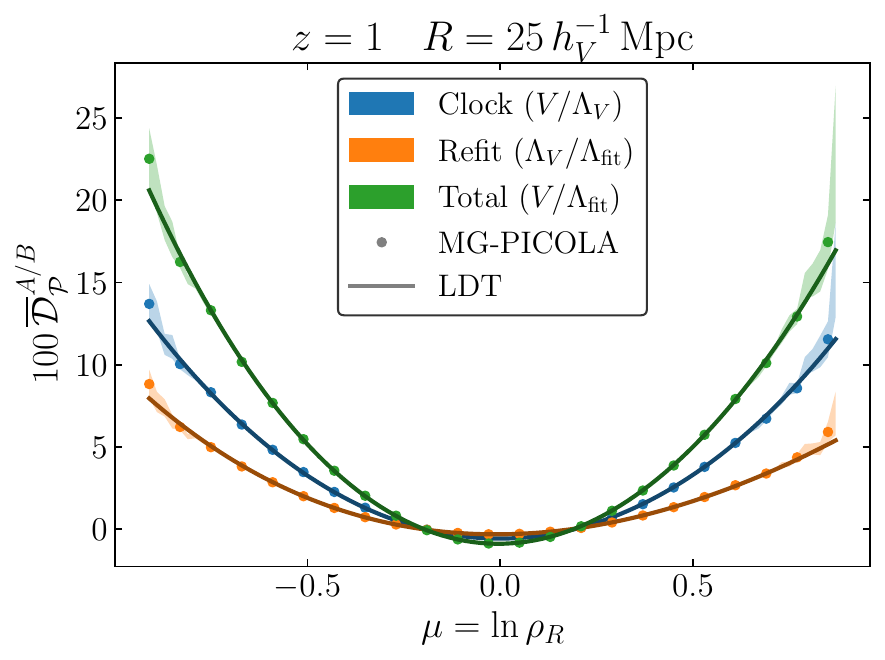}\hfill
 \includegraphics[width=0.24\textwidth]{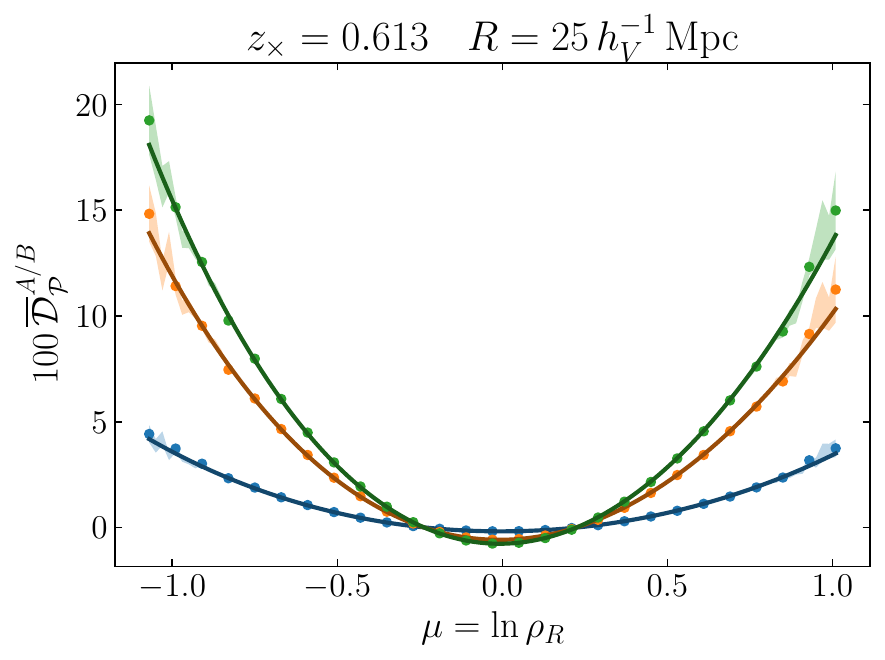}\hfill
 \includegraphics[width=0.24\textwidth]{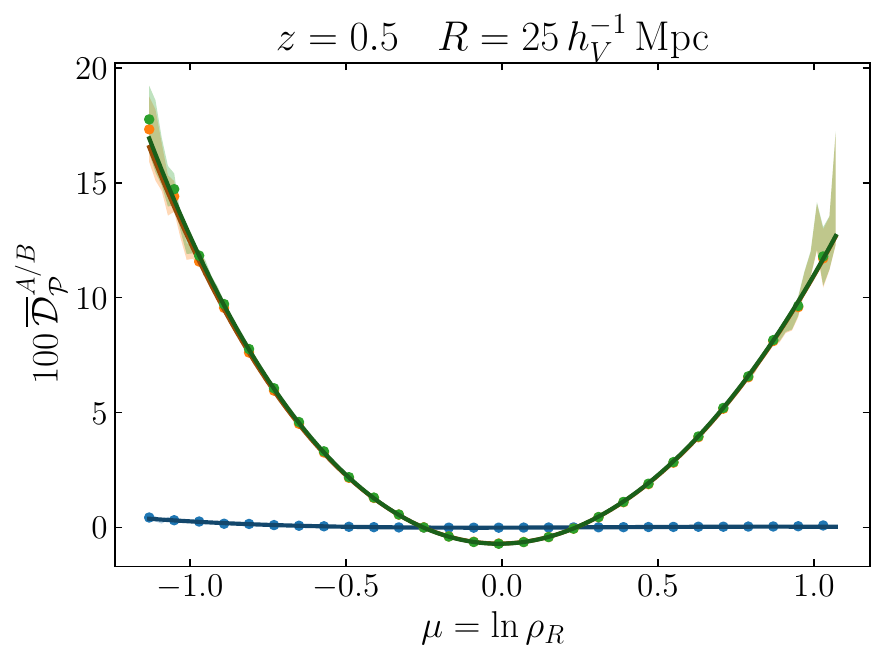}\hfill
 \includegraphics[width=0.24\textwidth]{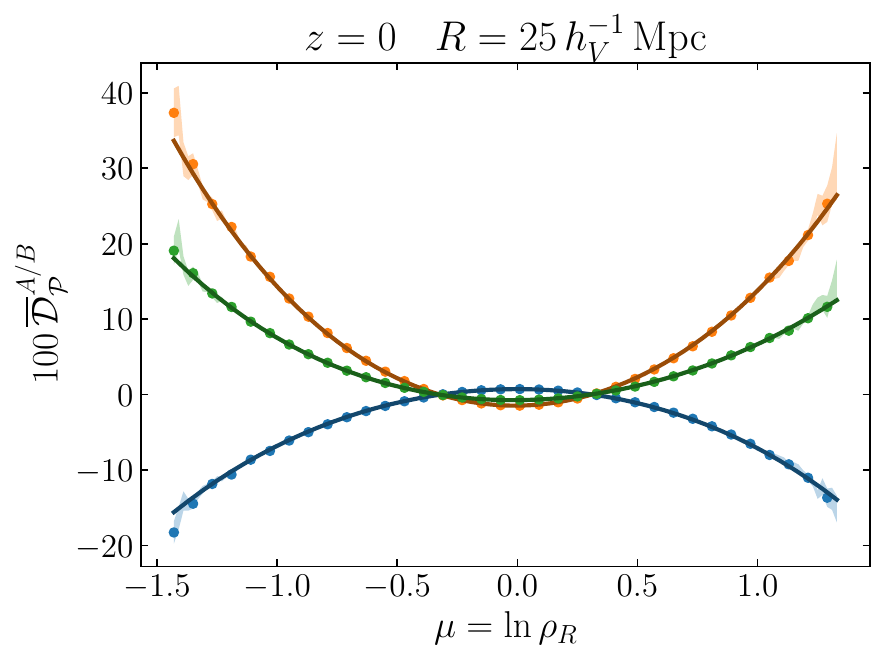}
 \caption{Same as Fig.~\ref{fig:triplet-pdf-v9e}, for
 $R=15$, $20$ and $25\,h_V^{-1}\,\mathrm{Mpc}$ (top to bottom; $22.53$,
 $30.04$ and $37.56\,\mathrm{Mpc}$), at $z=1$, $z_\times$, $0.5$ and $0$
 (left to right).}
 \label{fig:pdf-radius}
\end{figure*}

Figure~\ref{fig:pdf-radius} shows the PDF responses and LDT predictions
for $R=15$, $20$ and $25\,h_V^{-1}\,{\rm Mpc}$ ($22.53$, $30.04$ and
$37.56\,{\rm Mpc}$), in the same format as Fig.~\ref{fig:triplet-pdf-v9e}.
The behaviour is the same as at $R=10\,h_V^{-1}\,{\rm Mpc}$, while the
plotted density range shrinks and the LDT residual decreases as the
radius increases and the field becomes less nonlinear.

\section{PDF shapes after standardization}
\label{app:standardized-pdf}

Figure~\ref{fig:triplet-pdf-v9e} shows that the two cosmologies have different log-density PDFs at the same $\mu$. How much of this difference remains if we remove the differences in the mean and width? To answer this question, for each model $M$ we introduce the standardized log-density
\begin{equation}
\begin{aligned}
 y_M&=\frac{\mu-\langle\mu\rangle_M}{\sigma_{\mu,M}},\\
 \mathcal P_{y,M}(y)&=\sigma_{\mu,M}\,
 \mathcal P_{\mu,M}\!\left(\langle\mu\rangle_M+\sigma_{\mu,M}y\right).
\end{aligned}
\end{equation}
Here $\mu=\ln\rho_R$ and $\sigma_{\mu,M}=\sqrt{\Var_M(\mu)}$. The factor $\sigma_{\mu,M}$ in the PDF is the Jacobian $\dd\mu/\dd y$, which preserves probability under the change of variable. If the moments are those of the PDF being transformed, the new PDF has zero mean and unit variance. Two PDFs that differ only in mean and width will therefore coincide after this transformation. The same $y$ generally corresponds to different physical densities in the two models.

For the simulations, we calculate one mean and one standard deviation for each model by combining the cells of all 20 realizations. We use these same two numbers for every realization of that model. Let $p_{M,s,j}$ be the fraction of cells in realization $s$ in the standardized bin $[y_j^-,y_j^+]$. We reconstruct this fraction from the cumulative histogram in $\mu$, evaluated at the corresponding bin edges $\langle\mu\rangle_M+\sigma_{\mu,M}y_j^\pm$. As in Sec.~\ref{sec:numerics}, we first take the log ratio for each matched pair and then average over the pairs:
\begin{equation}
 100\overline{\mathcal D}_{\mathcal P_y}^{V/\Lambda_V}(y_j)
 =\frac{100}{20}\sum_{s=1}^{20}
 \ln\!\left[\frac{p_{V,s,j}}{p_{\Lambda_V,s,j}}\right].
\end{equation}
The common bin width cancels in this ratio. As a check, taking the log ratio of the two mean PDFs instead changes the result by at most $2.3\times10^{-8}$ in probability-weighted rms of the log ratio (before multiplication by 100) on the common central bins. For the error bands, we resample the 20 matched pairs $10^4$ times and recompute the means, variances and bin probabilities in each draw.

We treat each LDT PDF in the same way, but use its own theoretical mean $m_M^{\rm T}$ and standard deviation $\sigma_M^{\rm T}$, calculated over its full numerical support. The variance calibration and normalization of Sec.~\ref{sec:ldt} are unchanged. The probability in a bin of $y$ and the resulting predicted log ratio are
\begin{align}
 p_{M,j}^{\rm T}
 &=\int_{m_M^{\rm T}+\sigma_M^{\rm T}y_j^-}
 ^{m_M^{\rm T}+\sigma_M^{\rm T}y_j^+}
 \mathcal P_{\mu,M}^{\rm LDT}(\mu)\,\dd\mu,\\
 \mathcal D_{\mathcal P_y}^{\rm LDT}(y_j)
 &=\ln\!\left[\frac{p_{V,j}^{\rm T}}{p_{\Lambda_V,j}^{\rm T}}\right].
\end{align}
Neither the simulated nor the theoretical PDF is renormalized to the plotted interval in $y$. We do not fit the standardized shape.

\begin{figure*}[t]
\centering
\includegraphics[width=\textwidth]
{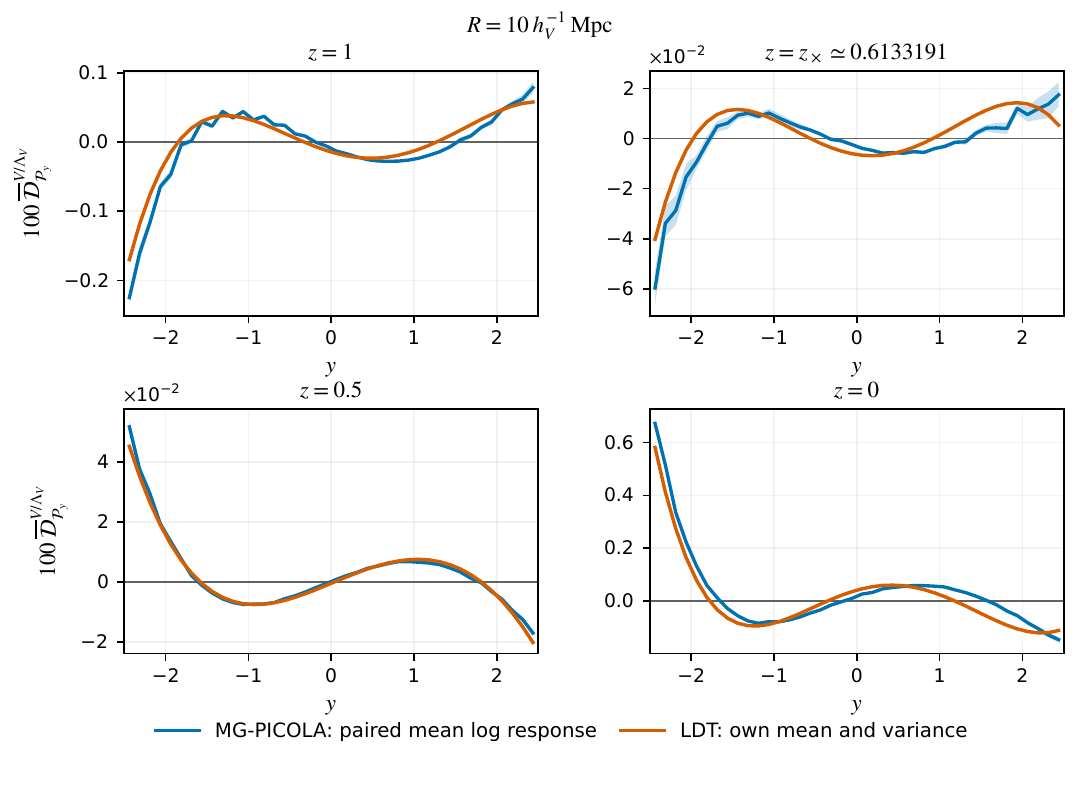}
\caption{Clock response of the standardized log-density PDF at $R=10\,h_V^{-1}\,\mathrm{Mpc}$ and $z=1$, $z_\times$, $0.5$ and $0$ (left to right, top to bottom). Blue curves show the mean log ratio from the original 20 phase-matched pairs, with pointwise 95\% paired-bootstrap bands. Orange curves show the LDT log ratio. Each model and each prediction is standardized using its own mean and standard deviation, with no new shape fit. Both curves use probabilities integrated over the same $y$ bins. The ordinate is the log ratio multiplied by 100.}
\label{fig:standardized-pdf}
\end{figure*}

Figure~\ref{fig:standardized-pdf} shows that the PDF differences become much smaller after standardization at $z=1$ and $z=0$. Near $z=0.5$, the difference at fixed $\mu$ was already small. The LDT prediction follows the main features of the remaining response. On the central bins used for the comparison at $R=10\,h_V^{-1}\,\mathrm{Mpc}$, the probability-weighted rms error of the predicted log ratio is between $5.8\times10^{-6}$ and $2.6\times10^{-4}$, corresponding approximately to $0.0006\%$--$0.026\%$ in the PDF ratio. These errors are larger than the tested numerical integration errors of LDT. The displayed bands account for scatter between realizations, but not for uncertainties in reconstructing the histograms or in the COLA dynamics.

\paragraph*{Skewness and the PDF maximum.}
The response in Fig.~\ref{fig:standardized-pdf} resembles the change expected from a small difference in skewness. To test this interpretation, we measure the skewness of the log-density,
\begin{equation}
 \gamma_{1,\mu,M}=\frac{\langle(\mu-\langle\mu\rangle_M)^3\rangle_M}
 {\sigma_{\mu,M}^3}=\langle y_M^3\rangle_M .
\end{equation}
For this measurement we rerun the simulations with the original settings and seeds, and calculate the moments directly on the CIC-deconvolved fields at $R=10\,h_V^{-1}\,\mathrm{Mpc}$. We combine all cells of the 20 realizations of each model, with no central-bin cut or shot-noise subtraction. All cells in these fields have positive, finite density.
We also calculate the skewness of the corresponding LDT PDFs, keeping their original variance calibration.

\begin{table}[t]
\centering
\caption{Difference in log-density skewness at $R=10\,h_V^{-1}\,\mathrm{Mpc}$. All numerical entries other than $z$ give $10^3\Delta\gamma_{1,\mu}$, where $\Delta\gamma_{1,\mu}=\gamma_{1,\mu,V}-\gamma_{1,\mu,\Lambda_V}$. The pointwise 95\% intervals are obtained by resampling the 20 matched pairs $10^4$ times and recomputing the moments in each draw. LDT is calibrated to the measured variance, not to the skewness.}
\label{tab:clock-r10-logdensity-skewness}
\begingroup
\setlength{\tabcolsep}{3pt}
\renewcommand{\arraystretch}{1.15}
\begin{tabular}{@{}crcr@{}}
\toprule
$z$ & Simulation & 95\% interval & LDT \\
\midrule
$1$         & $+0.79231$ & $[+0.79102,+0.79368]$ & $+0.51813$ \\
$z_\times$ & $+0.13303$ & $[+0.13261,+0.13349]$ & $+0.02576$ \\
$0.5$      & $-0.18946$ & $[-0.18953,-0.18939]$ & $-0.19916$ \\
$0$        & $-1.52852$ & $[-1.53092,-1.52614]$ & $-0.94908$ \\
\bottomrule
\end{tabular}
\endgroup
\end{table}

Table~\ref{tab:clock-r10-logdensity-skewness} compares the simulated difference in skewness between $V$ and $\Lambda_V$ with the LDT prediction. LDT reproduces its sign at all four epochs and gives the closest agreement in magnitude at $z=0.5$. The intervals measure sampling variation in COLA, not the accuracy of its dynamics. As a numerical check, reconstructing the skewness difference from fine histograms ($\Delta\mu=0.002$), including the measured moments of cells outside the histogram range, changes it by at most $1.4\times10^{-7}$.

Can the difference in skewness explain a shift in the PDF maximum? Retaining only the leading skewness term in a near-Gaussian expansion around $\varphi(y)=\e^{-y^2/2}/\sqrt{2\pi}$ gives
\begin{equation}
 \mathcal P_y(y)\simeq\varphi(y)\left[1+\frac{\gamma_1}{6}(y^3-3y)\right].
\end{equation}
At this order the maximum is at $y_{\rm mode}\simeq-\gamma_1/2$: a more positively skewed PDF peaks at smaller $y$. Across the four epochs and both models, LDT predicts smaller skewness ($0.15$--$0.21$) than COLA ($0.19$--$0.29$). Accordingly, the maximum of each LDT PDF lies at larger $y$ than its simulated counterpart, by about $0.02$--$0.03$. However, the much smaller difference in peak position between the two cosmologies is sensitive to the histogram binning and the method used to locate the peak. We cannot reliably determine that difference. Moreover, the maximum of a PDF and the extrema of a PDF ratio are different quantities.

The same near-Gaussian approximation gives a leading log-ratio contribution proportional to $(\gamma_{1,\mu,V}-\gamma_{1,\mu,\Lambda_V})(y^3-3y)/6$. To see how much of the discrepancy in Fig.~\ref{fig:standardized-pdf} can be accounted for by skewness, we subtract
\begin{equation}
 \frac{\Delta\gamma_{1,\mu}^{\rm sim}-\Delta\gamma_{1,\mu}^{\rm LDT}}{6}(y^3-3y)
\end{equation}
from the simulation-minus-LDT log ratio, averaging this term within each bin with a Gaussian weight. This is a diagnostic, with no fit to the response. At $z=1$, the rms discrepancy decreases from $9.14\times10^{-5}$ to $7.10\times10^{-5}$; at $z=0$ it is almost unchanged ($2.60\times10^{-4}$ in log-ratio units). Thus the leading skewness correction reduces the discrepancy at some epochs, but does not reproduce the full response difference. This test does not identify which higher moments or numerical approximations are responsible for the remainder.

\bibliography{f42_ldt}

\end{document}